\documentclass[preprint]{aastex}
\slugcomment{Accepted by ApJ}

\newcommand{\beq}{\begin{equation}}
\newcommand{\eeq}{\end{equation}}
\newcommand{\beqn}{\begin{eqnarray}}
\newcommand{\eeqn}{\end{eqnarray}}

\begin{document}

\shorttitle{Fully General Relativistic Simulations of Core-Collapse Supernovae}
\shortauthors{Kuroda, Kotake \& Takiwaki}

\title{Fully General Relativistic Simulations of Core-Collapse Supernovae with 
 An Approximate Neutrino Transport}

\author{Takami Kuroda\altaffilmark{1}, Kei Kotake\altaffilmark{1,2} and Tomoya Takiwaki\altaffilmark{2}}

\affil{$^1$Division of Theoretical Astronomy, National Astronomical Observatory of Japan, 
2-21-1, Osawa, Mitaka, Tokyo, 181-8588, Japan}
\affil{$^2$Center for Computational Astrophysics, National Astronomical Observatory of Japan, 2-21-1, Osawa, Mitaka, Tokyo, 181-8588, Japan}

\begin{abstract}
We present results from the first generation of multi-dimensional hydrodynamic core-collapse simulations in full general relativity (GR) that include an approximate treatment of neutrino transport. Using a M1 closure scheme with an analytic variable Eddington factor, we solve the energy-independent set of radiation energy and momentum based on the Thorne's momentum formalism. Our newly developed code is designed to evolve the Einstein field equation together with the GR radiation hydrodynamic equations. We follow the dynamics starting from the onset of gravitational core-collapse of a 15 $M_{\odot}$ star, through bounce, up to about 100 ms postbounce in this study. By computing four models that differ according to 1D to 3D and by switching from special relativistic (SR) to GR hydrodynamics, we study how the spacial multi-dimensionality and GR would affect the dynamics in the early postbounce phase. Our 3D results support the anticipation in previous 1D results that the neutrino luminosity and average neutrino energy of any neutrino flavor in the postbounce phase increase when switching from SR to GR hydrodynamics. This is because the deeper gravitational well of GR produces more compact core structures, and thus hotter neutrino spheres at smaller radii. By analyzing the residency timescale to the neutrino-heating timescale in the gain region, we show that the criterion to initiate neutrino-driven explosions can be most easily satisfied in 3D models, irrespective of SR or GR hydrodynamics. Our results suggest that the combination of GR and 3D hydrodynamics provides the most favorable condition to drive a robust neutrino-driven explosion.
 \end{abstract}


\section{Introduction}
\label{sec:Introduction}
 Core-collapse supernova simulations have been counted as
 one of the most challenging subjects in computational astrophysics. 
 The four fundamental forces of nature are all at play; 
 the collapsing iron core bounces due to strong interactions;
 weak interactions determine the energy and lepton number loss in the core 
via the transport of neutrinos
; electromagnetic interactions determine the 
properties of the stellar gas; general relativity plays an important role 
 due to the compactness of the proto-neutron star and also due to high velocities of the collapsing material outside.
 Naturally, such physical richness ranging from 
 a microphysical scale (i.e. femto-meter scale)
 of strong/weak interactions to a macrophysical 
scale of stellar explosions has long attracted the interest of researchers,
  necessitating a world-wide, multi-disciplinary collaboration to clarify
 the theory of massive stellar core-collapse and the formation 
 mechanisms of compact objects.

Ever since the first numerical simulation of such events \citep{Colgate66}, 
the neutrino-heating mechanism \citep{Wilson85,Bethe85},
 in which a stalled bounce shock could be revived via neutrino absorption 
 on a timescale of several hundred milliseconds after bounce,
 has been the working hypothesis of supernova theorists for these $\sim$ 45 years.
However, the simplest, spherically-symmetric (1D) form of this mechanism
 fails to blow up canonical massive stars \citep{Rampp00,Liebendorfer01,thom03,Sumiyoshi05}. 
  Pushed by mounting supernova observations of the blast morphology
 (e.g., \citet{wang01,Maeda08,Tanaka09}, and references therein), 
 it is now almost certain that the breaking of the spherical symmetry holds 
 the key to solve the supernova problem.
 So far a number of multidimensional (multi-D) hydrodynamic simulations have shown
 that hydrodynamic motions associated with convective overturn (e.g., 
\citet{Herant92,Burrows95,Janka96,frye02,fryer04a}) and the 
Standing-Accretion-Shock-Instability (SASI, e.g., 
\citet{Blondin03,Scheck04,scheck06,Ohnishi06,ohnishi07,Foglizzo06,Iwakami08,iwakami2,
  Murphy08,rodrigo09,rodrigo09_2}, and
 references therein) can help the onset of the neutrino-driven explosion. 
 
In fact, the neutrino-driven explosions have been obtained
 in the following first-principle two-(2D) and three-(3D) dimensional simulations 
in which the spectral neutrino transport is solved
 at various levels of approximations (e.g.,
 \cite{bernhard11,Ott11,Kotake11} for recent status reports).
 Among them are the 2D neutrino-radiation-hydrodynamic  
simulations by \citet{Buras06a,Buras06b,Marek09} who 
included one of the best available neutrino transfer approximations by the 
 ray-by-ray variable Eddington factor method, by \citet{bruenn} who included a
 ray-by-ray multi-group flux-limited diffusion (MGFLD) transport with 
the best available weak interactions, 
 and by \citet{Suwa10,Suwa11} who employed a ray-by-ray 
isotropic diffusion source approximation (IDSA) \citep{idsa} 
with a reduced set of weak interactions. 
  By extending the 2D modules in \citet{Suwa10}, \citet{Takiwaki11} 
recently reported neutrino-driven explosion models in 3D
 for an 11.2 $M_{\odot}$ star.
 They pointed out whether 3D effects would help
 explosions or not is sensitive to the employed numerical resolutions (see also 
\cite{Hanke11,Nordhaus10}). They argued that future peta- and exa-scale resources 
are at least needed to draw a robust conclusion of the 3D effects.

In addition to the 3D effects, impacts of general relativity (GR) on
 the neutrino-driven mechanism stand out among the biggest open questions  
  in the supernova theory. It should be remembered that using newly derived 
Einstein equations \citep{Misner64}, the consideration of GR was standard in the 
pioneering era of supernova simulations (e.g., \citet{May66}).
One year after \citet{Colgate66}, \cite{Schwarz67} reported the 
first fully GR simulation of stellar collapse 
to study the supernova mechanism, who implemented a 
gray transport of neutrino diffusion in the 
1D GR hydrodynamics\footnote{Citing from his paper, 
``{\it In this calculation, the neutrino
 luminosity of the core is found to be $10^{54}$ erg/s, or 1/2 a solar rest mass
 per second !! .... This is the mechanism which the supernova explodes".} The 
 neutrino luminosity rarely becomes so high in the modern simulations, but it is 
 surprising that the potential impact of GR on the neutrino-heating mechanism
 was already indicated in the very first GR simulation.}.
  Using GR Boltzmann equations derived by \citet{Lindquist66},
\citet{Wilson71} developed a 1D GR-radiation-hydrodynamic code 
including a more realistic (at the time) description
 of the collisional term than the one in \cite{Schwarz67}. 
 By performing 1D GR hydrodynamic simulations that included 
a leakage scheme for neutrino cooling, hydrodynamical properties 
 up to the prompt shock stagnation were studied in detail 
\citep{VanRiper79,VanRiper81,VanRiper82}. These pioneering studies, albeit 
 using a much simplified neutrino physics than today, did provide a 
 bottom-line of our current understanding of the supernova mechanism
(see \citet{Bruenn01} for a complete list of references for the early GR studies).
 In the middle of the 1980s, 
 \citet{Bruenn85} developed a code that coupled 1D GR hydrodynamics to 
 the MGFLD transport of order $(v/c)$ including the so-called standard
 set of neutrino interactions.
Since the late 1990s, 
 the ultimate 1D simulations, in which the GR Boltzmann transport
 is coupled to 1D GR hydrodynamics, have been made feasible by Sumiyoshi-Yamada et al. \citep{Yamada97,Yamada99,Sumiyoshi05,Sumiyoshi07}\footnote{Very recently, they reported
 their success to develop the first multi-angle, multi-energy neutrino transport
 code in 3D \citep{Sumi11}.}
 and by Liebend\"orfer-Mezzacappa-Bruenn et al. 
\citep{Mezzacappa89,Bruenn01,Liebendorfer01,Matthias01,matthias04} (and by their collaborators).

Among them, \citet{Bruenn01} presented evidence that 
average neutrino energy of any neutrino flavor during the shock reheating phase
 increase when switching from Newtonian to GR hydrodynamics. 
 They also pointed out that the increase is larger in 
 magnitude compared to the decrease due to redshift effects and gravitational
 time dilation. By employing the currently best available weak interactions,
 \citet{Lentz11} very recently reported the update of \citet{Bruenn01}.
  They showed that the omission of observer corrections in the transport equation 
 particularly does harm to drive the neutrino-driven explosions. 
  In these full-fledged 1D simulations, a commonly observed disadvantageous aspect of GR 
 to drive neutrino-driven explosions is 
that the residency time of material
 in the gain region becomes shorter due to the stronger gravitational pull. As a result of these competing ingredients in the end, GR
  works disadvantageously to facilitate the neutrino-driven explosions 
in 1D. In fact, the maximum shock extent in the postbounce phase 
 is shown to be 20\% smaller when switching from Newtonian to GR hydrodynamics 
 (e.g., Figure 2 in \cite{Lentz11}). 

 Among the most up-to-date multi-D models with 
 spectral neutrino transport mentioned earlier, 
  GR effects are at best attempted to be modeled by replacing 
 the monopole term in the Newtonian potential with an effective 
Tolman-Oppenheimer-Volkov potential \citep{Buras06a,Buras06b,Marek09,bruenn}.
 A possible drawback of this prescription is that a conservation law for the 
 total energy
 cannot be guaranteed by adding an artificial term to the Poisson equation of 
 self-gravity. Since
 the energy reservoir of the supernova engines is the gravitational binding energy,
 any potential inaccuracies in the argument of gravity would be better eliminated.
There are a number of relativistic simulations of massive stellar collapse 
 in full GR (e.g., 2D \citep{Shibata05a} or 3D 
  \citep{Shibata05b,Ott2007}, and references therein) 
or using the conformally-flatness approximation (CFC) 
(e.g., \citet{Dimmelmeier02,Isa2009}).
 Although extensive attempts have been made to include
 microphysics such as by the $Y_e$ formula \citep{Matthias05} or 
 by a neutrino leakage scheme \citep{Sekiguchi10}, 
 the effects of neutrino heating have yet to be included in them, 
which is a main hindrance
 to study the GR effects on the multi-D neutrino-driven
 mechanism\footnote{Very recently, \citet{Bernhard12}
 reported explosions for $11.2$ and $15 M_{\odot}$ stars based on their 
 2D GR simulations in CFC
 with detailed neutrino transport \citep{Bernhard10} similar to \citet{Buras06a}.}.

 In this paper, we present a new fully GR code 
 for multi-D hydrodynamic supernova simulations in which an approximate neutrino 
transport is implemented. The code is a marriage of an adaptive-mesh-refinement (AMR),
 conservative 3D GR magnetohydrodynamic (MHD) code developed by \citet{Kuroda10}, 
and the approximate neutrino transport code that we newly develop in this work. 
The spacetime treatment in our full GR code
is based on the Baumgarte-Shapiro-Shibata-Nakamura (BSSN) 
formalism \citep[see, e.g.,][]{Shibata95,Baumgarte99}.
 The hydrodynamics can be 
 solved either in full GR or in special relativity (SR), which allows us to 
 investigate the GR effects on the supernova dynamics. 
We solve the energy-independent set of radiation moments
 up to the first order and evaluate the second order momentum
 with an analytic variable Eddington factor
(the so-called M1 closure scheme \citep{Levermore84}).
 This part is based on the partial implementation of the Thorne's momentum formalism,
  which is recently extended by \citet{Shibata11} in 
a more suitable manner applicable to the neutrino transport problem.
 Similar to the isotropic diffusion source approximation (IDSA scheme \citep{idsa}), we conceptually
 divide the neutrinos into two parts, which are ``trapped" and ``free-streaming" neutrinos.
 By doing so, we model the source terms of the transport equations to be expressed 
 in a simplified manner with the use of a multi-flavor
 neutrino leakage scheme (e.g., \citet{Rosswog03}).
Our newly developed code is designed to evolve the Einstein field
 equation together with the GR radiation hydrodynamic equations
 in a self-consistent manner while satisfying the Hamiltonian and momentum constraints. 
  An adaptive-mesh-refinement technique implemented in the 3D 
 code enables us to follow the dynamics starting from the onset of gravitational core-collapse of a 15 $M_{\odot}$ star, through bounce, 
up to about 100 ms postbounce in this study.
 For the 15 $M_{\odot}$ star,
the neutrino-driven explosions are expected to 
 take place later than $\sim 200$ ms postbounce at the earliest (e.g., \citet{bruenn,Marek09}). However it is computationally too expensive 
to follow such a long-term evolution in our full 3D GR simulations.
 Albeit limited to the rather early postbounce phase, we would 
 exploratory study possible GR effects in the multi-D neutrino-driven mechanism 
 by comparing 1D to 3D results
 and by switching from SR to GR hydrodynamics.

This paper is organized as follows: In section 2, after we introduce 
 the model concept of the approximate GR transport scheme, we summarize the governing 
 equations of hydrodynamics and neutrino transport in detail.
  The main results are presented in Section 3.
We summarize our results and discuss their implications in Section 4.
Note that geometrized units are used throughout Sections 2 to 3, i.e. both the speed 
 of light and the gravitational constant are set to unity: $G = c = 1$.
Greek indices run from 0 to 3, Latin indices from 1 to 3.

\section{Basic Equations for General Relativistic Neutrino-Radiation 
Hydrodynamics}
\label{sec:Formulation}

Our newly developed code consists of the three parts,
in which the evolution equations
 of metric, hydrodynamics, and neutrino radiation are solved, respectively.
 As will be mentioned, each of them 
 is solved in an operator-splitting manner, but the system evolves self-consistently 
as a whole satisfying the Hamiltonian and momentum constraints.
 Before going into details,  we shortly describe the bottom-line 
how to add radiation to GR hydrodynamics. 

The starting-point is the conservation of the total energy (fluid + radiation), 
\begin{equation}
\nabla_{\alpha} T_{\rm (total)}^{\alpha\beta} = 
\nabla_{\alpha} T_{\rm (fluid)}^{\alpha\beta} + \nabla_{\alpha} T_{(\nu)}^{\alpha\beta} = 0,
\label{eq1}
\end{equation}
 where $T_{\rm (total)}^{\alpha\beta}$, $T_{\rm (fluid)}^{\alpha\beta}$, 
and $T_{(\nu)}^{\alpha\beta}$ is the stress-energy 
tensor of the total energy, fluid, and neutrino radiation, respectively.
 Then Equation (\ref{eq1}) can be decomposed as,
\begin{equation}
\nabla_{\alpha} {{T}_{\rm (fluid)}^{\alpha\beta}} = - Q^{\beta},
\label{eq5}
\end{equation}
and 
\begin{equation}
\nabla_{\alpha} T_{(\nu)}^{\alpha\beta} =  Q^{\beta},
\label{eq6}
\end{equation}
where $Q^{\beta}$ represents the source terms 
 that describe the exchange of 
 energy and momentum between fluid and radiation. 
If $Q^{\beta}$ would be given, it is rather straightforward to evolve Equation
 (\ref{eq5}) following standard procedures in numerical relativity.
  Accordingly, what we focus on in this section is how to determine the  
 source terms $Q^{\beta}$ and the evolution equations of neutrinos
 (i.e. the concrete form of the left-hand-side of Equation (\ref{eq6})).
 In doing so, there will appear many terms related to 
 GR such as $e^{6\phi},\,\partial 
_i \beta^{j}$.. etc. So, after we briefly summarize the BSSN formalism 
in the next section, we first present the transport equations
 in section \ref{neu_rad} and then the source terms 
in section \ref{source}.

\subsection{Metric equations}
\label{sec:metric}
We write the spacetime metric in the standard (3+1) form:
\begin{eqnarray}
\label{eq:line element}
ds^2=-\alpha^2dt^2+\gamma_{ij}(dx^i+\beta^idt)(dx^j+\beta^jdt),
\end{eqnarray}
where $\alpha$, $\beta^i$, and $\gamma_{ij}$ are the lapse, shift, 
 and spatial metric, 
respectively. The extrinsic curvature $K_{ij}$ is
defined by
\begin{equation}
\label{Kij}
(\partial_t - {\mathcal{L}}_{\beta})\gamma_{ij} = -2\alpha K_{ij},
\end{equation}
where ${\mathcal{L}}_{\beta}$ is the Lie derivative with respect to
$\beta^i$. The evolution of $\gamma_{ij}$ and $K_{ij}$ is governed 
by the Einstein equation $G_{\mu\nu} = 8 \pi T_{\mu\nu\,{\rm(total)}}$, where 
$G_{\mu \nu}$ is the Einstein tensor and ${T_{\mu \nu}}_{\rm (total)}$ is the 
 total stress-energy tensor (e.g., Equation (\ref{eq1})). 

We evolve $\gamma_{ij}$ and $K_{ij}$
using the BSSN 
formulation \citep{Baumgarte99,Shibata05b,Duez06}, in which the fundamental variables are
\begin{eqnarray}
  \phi &\equiv& \frac{1}{12}\ln[\det(\gamma_{ij})]\ , \\
  \tilde\gamma_{ij} &\equiv& e^{-4\phi}\gamma_{ij}\ , \\
  K &\equiv& \gamma^{ij}K_{ij}\ , \\
  \tilde A_{ij} &\equiv& e^{-4\phi}(K_{ij} - \frac{1}{3}\gamma_{ij}K)\ , \\
  \tilde\Gamma^i &\equiv& -\tilde\gamma^{ij}{}_{,j}\ .
\end{eqnarray}

 The Einstein equation gives rise to the evolution equations for the BSSN variables as,
\begin{eqnarray}
\label{eq:BSSN1}
(\partial_t-\mathcal{L}_\beta)\tilde\gamma_{ij}&=&-2\alpha\tilde A_{ij} \\
\label{eq:BSSN2}
(\partial_t-\mathcal{L}_\beta)\phi&=&- \frac{1}{6}\alpha K \\
\label{eq:BSSN3}
(\partial_t-\mathcal{L}_\beta)\tilde A_{ij}&=&e^{-4\phi}\left[ \alpha (R_{ij} -8\pi \gamma_{i\mu}\gamma_{j\nu}T^{\mu\nu}_{\rm (total)}-D_iD_j \alpha\right]^{\rm trf}+\alpha(K\tilde A_{ij}-2\tilde A_{ik}\tilde \gamma ^{kl}\tilde A_{jl})\nonumber \\ \\
\label{eq:BSSN4}
(\partial_t-\mathcal{L}_\beta)K&=&-\Delta \alpha +\alpha (\tilde A_{ij}\tilde A^{ij}+K^2/3)+4\pi \alpha (n_\mu n_\nu T^{\mu\nu}_{\rm (total)}+\gamma^{ij}\gamma_{i\mu}\gamma_{j\nu}
T^{\mu\nu}_{\rm (total)}) \\
\label{eq:BSSN5}
(\partial_t-\beta^k\partial_k)\tilde\Gamma^i&=&16\pi\tilde\gamma^{ij}\gamma_{i\mu}n_\nu 
T^{\mu\nu}_{\rm (total)} \nonumber\\
&&-2\alpha(\frac{2}{3}\tilde\gamma^{ij}K_{,j}-6\tilde A^{ij}\phi_{,j}-\tilde\Gamma^i_{jk}\tilde A^{jk})\nonumber\\
&&+\tilde\gamma^{jk}\beta^i_{,jk}+\frac{1}{3}\tilde\gamma^{ij}\beta^k_{,kj}-\tilde\Gamma^{j}\beta^i_{,j}
+\frac{2}{3}\tilde\Gamma^{i}\beta^j_{,j}+\beta^j\tilde\Gamma^i_{,j}-2\tilde A^{ij}\alpha_{,j},
\end{eqnarray}
where $D$ denotes covariant derivative operator associated 
with $\gamma_{ij}$, $\Delta=D^iD_i$,``trf'' denotes the trace-free operator, 
$n_\mu=(-\alpha,0)$ is the time-like unit vector normal to 
 the $t=$ constant time slices. In Equation (\ref{eq:BSSN3}), the
 explicit form of $D_iD_j \alpha$ reads
\begin{eqnarray}
\label{eq:DDalpha}
D_i D_j \alpha&=&\partial_i\partial_j \alpha-\Gamma^k_{ij}\partial_k\alpha \nonumber\\
&=&\partial_i\partial_j \alpha
-\left[\tilde\Gamma^k_{ij}+2\left(\delta_j^k\partial_i\phi +\delta_i^k\partial_j\phi -\tilde\gamma_{ij}\tilde\gamma^{kl}\partial_l\phi \right)\right]\partial_k\alpha.
\end{eqnarray}
Following  \citet{Alcubierre01}, the gauge is specified by the 1+log lapse, 
\begin{equation}
\label{eq:1+log}
\partial_t\alpha=\beta^i\partial_i\alpha-2\alpha K,
\end{equation}
and by the Gamma-driver-shift,
\begin{equation}
\label{eq:GammaDriver}
\partial_t\beta^i=k\partial_t \tilde\Gamma^i,
\end{equation}
here we chose $k=1$. For further information with code verification
 of the metric solver, see \citet{Kuroda10}\footnote{In \cite{Kuroda10}, an auxiliary variable 
$F_i\equiv\delta^{jk}\partial_k\tilde \gamma_{ij}$ was evolved 
 instead of $\tilde\Gamma^i$.}.
In addition, during calculations, we enforce following algebraic constraints every after the
time updating to satisfy $\tilde\gamma=1$ and $\tilde {A^i}_i=0$ \citep{Zlochower05,Etienne08}.
\begin{eqnarray}
\tilde\gamma_{ij}&\rightarrow&\tilde\gamma_{ij}\tilde\gamma^{-1/3}\\
\tilde A_{ij}&\rightarrow&\tilde A_{ij}-\frac{\tilde\gamma_{ij}}{3}\tilde {A^k}_k
\end{eqnarray}

Having summarized the bottom-line of the BSSN formalism, we are now
 ready to discuss the transport equations. 

\subsection{Neutrino Transport Equations}
\label{neu_rad}

To determine the transport equations in GR,
we follow the truncated momentum formalism \citep{Thorne81}, which is
recently extended by 
 \citet{Shibata11} in a suitable form for
 the neutrino transport problem. The starting point
 is to define the radiation stress-energy tensor as,
\begin{eqnarray}
{T_{(\nu)}}^{\alpha\beta} \equiv E_{(\nu)} n^\alpha n^\beta+ {F_{(\nu)}}^\alpha n^\beta+
 {F_{(\nu)}}^\beta n^\alpha +
{P_{(\nu)}}^{\alpha\beta},
\label{t_lab}
\end{eqnarray}
where $E_{(\nu)}$, $F_{(\nu)}$, and $P_{(\nu)}$, 
is the radiation energy, flux, pressure measured in the laboratory frame, respectively.
 Conversely, $E_{(\nu)}$, $F_{(\nu)}^{~\alpha}$, and 
$P_{(\nu)}^{~\alpha\beta}$ are given by the stress-energy tensor as,
\beqn
E_{(\nu)}=T_{(\nu)}^{~\alpha\beta}n_{\alpha}n_{\beta},~~~
F_{(\nu)}^{~i}=-T_{(\nu)}^{~\alpha\beta}n_{\alpha}\gamma_{\beta}^{~i},~~~
P_{(\nu)}^{~ij}=T_{(\nu)}^{~\alpha\beta}
\gamma_{\alpha}^{~i}\gamma_{\beta}^{~j}.
\label{e_def}
\eeqn
 In the following, radiation variables are all defined in 
 the laboratory frame unless otherwise stated.

According to \citet{Shibata11}, the evolution equations of 
 radiation energy and radiation flux in Equation (\ref{t_lab}) can be written as 
\begin{equation}
\partial_t (e^{6\phi}E_{(\nu)})+\partial_i [e^{6\phi}(\alpha F_{(\nu)}^i-\beta^i E_{(\nu)})] =
e^{6\phi}(\alpha P^{ij}K_{ij}-F_{(\nu)}^i\partial_i \alpha-\alpha Q^\mu n_\mu),
\label{rad1}
\end{equation}
and 
\begin{equation}
\partial_t (e^{6\phi}{F_{(\nu)}}_i)+\partial_j [e^{6\phi}(\alpha 
{P_{(\nu)}}_i^j-\beta^j {F_{(\nu)}}_i)] =
e^{6\phi}[-E_{(\nu)}\partial_i\alpha +{F_{(\nu)}}_j\partial_i \beta^j+(\alpha/2) 
P_{(\nu)}^{jk}\partial_i \gamma_{jk}+\alpha Q^\mu \gamma_{i\mu}],
\label{rad2}
\end{equation}
 where $Q^{\mu}$ denotes the source terms. For the three 
 radiation variables ($E_{(\nu)}$, $F^i_{(\nu)}$, $P^{ij}_{(\nu)}$) in
  Equations (\ref{rad1},\ref{rad2}), we have only two sets of
 the equation. Here we employ the so-called M1 
closure \citep{Levermore84}, in which the radiation pressure is related to 
  the radiation energy and flux by an analytical closure relation (i.e. 
$P_{(\nu)}(E_{(\nu)}, F_{(\nu)})$) as,
\begin{eqnarray}
{P_{(\nu)}}^{ij}=\frac{3\chi-1}{2}P^{ij}_{\rm thin}+\frac{3(1-\chi)}{2}P^{ij}_{\rm 
thick},
\label{neu_p}
\end{eqnarray}
 where $\chi$ represents the variable Eddington factor,
$P^{ij}_{\rm thin}$ and 
$P^{ij}_{\rm thick}$ corresponds to the radiation pressure in the optically thin and 
 thick limit, respectively.

For the variable Eddington factor $\chi$, we employ the one proposed by 
\citet{Levermore84},
\begin{eqnarray}
\label{eq:Livermore_chi}
\chi&=&\frac{3+4\bar{F}^2}{5+2\sqrt{4-3\bar{F}^2}},\\
\bar{F}^2&\equiv&\frac{F^iF_i}{E^2}.
\end{eqnarray}
It can be readily checked that 
in the optically thick limit, $\chi \rightarrow 1/3$ because $F^i \rightarrow 0$,
 then $P^{ij} \rightarrow P^{ij}_{\rm thick}$, while in 
 the optically thin limit, $\chi \rightarrow 1$ because 
$\bar{F}^2 \rightarrow 1$, then $P^{ij} \rightarrow P^{ij}_{\rm thin}$.

Following \citet{Audit02,Shibata11}, the following forms of
$P^{ij}_{\rm thin}$ and $P^{ij}_{\rm thick}$ are adopted,
\begin{eqnarray}
\label{eq:PijThin}
P^{ij}_{\rm thin}=E\frac{F^iF^j}{F_kF^k},
\end{eqnarray}
and 
\begin{eqnarray}
\label{eq:PijThick}
P^{ij}_{\rm thick}=\mathcal{J}\frac{\gamma^{ij}+4\gamma^{ik}\gamma^{jl}u_k u_l}{3}+\gamma^{jk}\mathcal{H}^i u_k
+\gamma^{ik}\mathcal{H}^j u_k,
\end{eqnarray}
 respectively. By this choice, the radiation flux naturally changes
with radius ($r$) as $\sim 1/r^{2}$ in the low opacity regime (e.g., Appendix \ref{appC}).
 This may sound quite straightforward, but it is one of the most
  important issue for the purpose of this work, because the radiation
   neutrino flux in the
  semi-transparent regions holds the key to the success or failure of
  the neutrino-heating mechanism. $\mathcal{J, \, H}$ in Equation (\ref{eq:PijThick}) 
 denotes the Eddington moments in the comoving frame, which are
 related to those in the laboratory frame as,
\beq
\mathcal{J} = E_{(\nu)}W^2 - 2 W {F_{(\nu)}}^i u_i + {P_{(\nu)}}^{ij} u_i u_j,
\label{j1}
\eeq
and
\beq
\mathcal{H}^{\alpha} = ({E_{(\nu)}}W - {F_{(\nu)}}^{i} u_i)(n^{\alpha} -
 W u^{\alpha}) + W h^{\alpha}_{\beta} {F_{(\nu)}}^{\beta} - h^{\alpha}_i u_j 
{P_{(\nu)}}^{ij},
\label{j2}
\eeq
 where $W =\alpha u^0$ is the Lorentz factor, $h_{\alpha\beta}$ is the
 projection
 operator defined by 
\beqn
h_{\alpha\beta} \equiv g_{\alpha\beta}+u_{\alpha}u_{\beta}. 
\eeqn
 Having summarized the closed set of the two-moment
 transport equations, we are now going to discuss
 the source terms (;$Q^{\mu}$)  in the next section.


\subsection{Source terms}\label{source}

 To model the source terms, 
 we follow the idea of the IDSA scheme \citep{idsa}, in which 
 neutrinos are conceptually divided into two parts, which are ``trapped" and 
``free-streaming" neutrinos, respectively. We also utilize a methodology of multi-flavor
 neutrino leakage scheme (e.g., \citet{Rosswog03}) to 
simplify the description of the source terms.

%

\begin{figure}[htpb]
\begin{center}
\includegraphics[width=100mm,angle=0.]{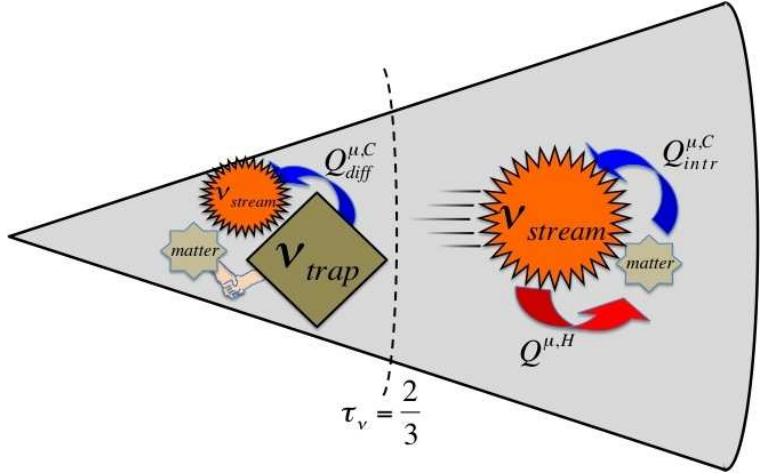}
\end{center}
  \caption{
A schematic illustration how to model the source terms in 
  our transport scheme.
In this figure, trapped and streaming neutrinos are represented by ``$\nu_{\rm trap}$'' 
and ``$\nu_{\rm stream}$", respectively.  $Q^{\mu,C}_{diff}$, $Q^{\mu,C}_{intr}$, 
 and $Q^{\mu,H}$ denotes the coupling terms between them (see text for more details).}
\label{pic}
\end{figure}

Figure \ref{pic} illustrates the concept how to estimate the source terms.
 To describe the neutrino-matter coupling, 
we need to ask at least three actors,
 namely ``matter", ``trapped neutrino", and ``streaming neutrino",
 to appear in the playground of the supernova core.
 The neutrino sphere is an important quantity to describe the relationship between them.
 The position of the neutrino sphere\footnote{here defined for the average neutrino 
energy for simplicity,} at which the neutrino optical depth comes close to unity,
 is indicated by $\tau_{\nu} = 2/3$ in Figure \ref{pic}. 
 The trapped neutrinos are always coupled with matter 
(through $\beta$-equilibrium) and their temperature is the same as that of matter.
On the other hand, temperature of the free-streaming neutrinos cannot be 
 determined locally owing to its decoupling nature from matter, 
 which was the reason that we have to solve the evolution equations.
 So these two represent a two extreme limit.

 In Figure \ref{pic}, trapped neutrinos are denoted by ``$\nu_{\rm trap}$" 
(in a diamond shape colored by grey),
 which is illustrated to shake hands with matter inside the 
 neutrino sphere (inside the region enclosed by $\tau_{\nu} = 2/3$).
 There the trapped neutrinos dominate over the streaming neutrinos 
(denoted by ``$\nu_{\rm stream}$"  (in a jaggy circle colored by orange) 
in Figure \ref{pic}), which is vice versa outside the  neutrino sphere. 
This is illustrated in such a way that
 ``$\nu_{\rm trap}$" is bigger than ``$\nu_{\rm stream}$" inside the 
 neutrino sphere, which is vice versa outside the neutrino sphere. 
For ``$\nu_{\rm stream}$"
 outside the neutrino sphere, the jaggy circle is drawn to have several
 tails, by which we intend
 to express that it can travel much more freely in the free-streaming regime.

\placefigure{pic}

In Figure \ref{pic}, the three actors are connected by thick arrows (in blue or red),
 each of them is labeled by $Q^{\mu,C}_{diff}$ (in blue),
 $Q^{\mu,C}_{intr}$ (in blue), 
 or $Q^{\mu,H}$ (in red), representing the couplings in-between.
The arrows colored by blue ($Q^{\mu,C}_{diff}$, $Q^{\mu,C}_{intr}$) represent 
 neutrino cooling, while the arrow in red ($Q^{\mu,H}$) does neutrino heating.
 The neutrino cooling means that energy is transferred from matter (or from trapped 
 neutrinos) to streaming neutrinos that carry the imparted energy 
away from the system. On the other hand, the neutrino heating proceeds
 by energy transfer from streaming neutrinos to 
 matter (see $Q^{\mu,H}$ in Figure \ref{pic}). Finally $Q^{\mu,C}_{diff}$ 
in Figure \ref{pic} represents
the cooling by neutrinos leaking out from opaque regions inside
 the neutrino sphere by diffusion.\footnote{Note inside the neutrino sphere,
 neutrino heating locally balances with neutrino cooling by weak interactions 
 due to $\beta$-equilibrium. So as a net, the diffusion-mediated cooling becomes
 dominant there.}

Looking at Figure \ref{pic} again, the source term of
 Equations (\ref{rad1},\ref{rad2}) can be 
readily defined as, 
\begin{equation}
Q^\mu \equiv  Q^{\mu,C} -Q^{\mu,H},
\label{eq:Qnu}
\end{equation}
 where each of the cooling ($Q^{\mu,C}$) and heating($Q^{\mu,H}$)
 term is calculated in the present scheme as,
\begin{eqnarray}
\label{eq:QmuC}
Q^{\mu,C}&=&\sum_{\nu\in\nu_e,\bar\nu_e,\nu_x}
\Bigl
[(1-e^{-\beta_\nu\tau_\nu})Q_{\nu,diff}+e^{-\beta_\nu\tau_\nu}Q_{\nu,intr}^C
\Bigr]u^\mu, \\
\label{eq:QmuH}
Q^{\mu,H}&=&\sum_{\nu\in\nu_e,\bar\nu_e}e^{-\beta_\nu\tau_\nu}\varepsilon_\nu^2\tilde\kappa_\nu\bigl(-\mathcal{J}_\nu u^\mu-\mathcal{H}_\nu^\mu\bigr).
\end{eqnarray}
 Before we go into details, we need to draw a caution that we introduced the concept 
 of the streaming and trapped neutrinos only for the sake of (better) explanation of  
our approximate treatment. Actually the sum of the trapped and streaming is 
transported by Equations (\ref{rad1},\ref{rad2}) with the source terms described 
above. We design the source terms in such a way to connect the heating/cooling terms 
 smoothly between the diffusion and free-streaming limit, 
 which is basically similar to the concept of the M1 closure relation.

The cooling term ($Q^{\mu,C}$) consists of 
 $Q_{\nu, diff}$ and $Q^{C}_{\nu, intr}$, which 
 accounts for neutrino cooling by diffusion out
 of the neutrino sphere and the one determined locally
 outside the neutrino sphere, respectively
 (e.g., Figure 1). Following \citet{VanRiper81}, 
the terms of $1- e^{-\beta_\nu\tau_\nu}$ and  
 $e^{-\beta_\nu\tau_\nu}$ appearing in Equation (\ref{eq:QmuC}) 
are introduced to smoothly connect the two quantities 
($Q_{\nu, diff}$ and $Q^{C}_{\nu, intr}$) for 
the semi-transparent regime. Here $\tau_{\nu}$ represents
 the optical depth of neutrinos and $\beta_{\nu}$ is a model parameter that we 
 determine by the comparison with a spectral neutrino transport
 calculation  (see Appendix \ref{sec:Neutrino diffusion terms}). 
 With these terms bridging the two limits, 
it is easy to see that $Q^{\mu,C}$ approaches
 to $Q_{\nu, diff}$ for the diffusion limit ($\tau_{\nu} \rightarrow \infty$), and
 it does to $Q^{C}_{\nu, intr}$ for the free-streaming limit 
($\tau_{\nu} \rightarrow 0$).

 Following the neutrino leakage scheme (e.g., \citet{Epstein81,
VanRiper81,Kotake03}), the diffusion cooling rate ($Q_{\nu,diff}$) 
can be given as 
\begin{eqnarray}
\label{diff1}
Q_{\nu,diff} [{\rm erg}/{\rm cm}^3/{\rm s}] &\equiv & 
\int{\frac{\varepsilon_\nu\,\, n_\nu(\varepsilon_\nu)}
{T^{diff}_\nu(\varepsilon_\nu)}d\varepsilon_\nu},   
\end{eqnarray}
  where the right-hand-side of Equation (\ref{diff1}) is simply 
 proportional to the leakage of the neutrino energy density
 $:\varepsilon_\nu\, n_\nu$ [erg/$\rm{cm}^3$] divided by the 
 diffusion timescale $T^{diff}_\nu$ [s]. More details to estimate these
 quantities as well as how to determine the neutrino sphere are summarized 
in Appendix A.
%
\begin{table}[htpb]
\begin{center}
\begin{tabular}{cc}
\hline\hline
Charged Current Interactions & \\
\hline
$n\nu_e\leftrightarrow e^-p$ &  \\
$p\bar{\nu}_e\leftrightarrow e^+n$  \\
$\nu_e A\leftrightarrow e^-A'$  \\
\hline
Neutral Current Interactions \\
\hline
$\nu p\leftrightarrow \nu p$ \\
$\nu n\leftrightarrow \nu n$  \\
$\nu A\leftrightarrow \nu A$  \\
\hline
\end{tabular}
\caption{The opacity set included in the present simulation. 
Note that $\nu$, in neutral current 
reactions, represents all species of neutrinos ($\nu_e,\bar\nu_e,\nu_x$) with $\nu_x$ 
 representing heavy-lepton neutrinos (i.e. 
$\nu_{\mu}, \nu_{\tau}$ and their anti-particles).}
\label{tb:NeutrinoInteraction}
\end{center}
\end{table}

Striving for simplification of our modeling, we include a reduced set of
 neutrino-matter interactions (e.g., Table 1). Regarding the charged-current
 interactions, emission and absorption of electron neutrinos by neutrons (the 
 first column in Table 1), emission and absorption of electron anti-neutrinos
 by proton (the second column), and emission and absorption of electron neutrinos 
 by heavy nuclei (the third column), are included.
 For the neutral-current interactions, elastic
 scattering of all neutrino flavors off nucleons (the fourth and fifth column in 
 Table 1), and the coherent elastic scattering (the sixth column) are included. 
 For the cross sections of these reactions, we employ the ones 
 summarized in \citet{Adam06} while omitting higher-order terms such as 
ion-ion correlations and weak magnetism for simplicity.

In computing the neutrino cooling rate ($Q_{\nu,intr}^C$), we furthermore consider 
 the contribution from pair neutrino annihilation $Q_{e^-e^+\rightarrow \nu\bar\nu}$ 
 \citep{Cooperstein86},
 plasmon decay $Q_{\gamma\rightarrow \nu\bar\nu}$ \citep{Ruffert96}, and 
 nucleon-nucleon bremsstrahlung $Q_{NN \rightarrow NN \nu\bar\nu}$ \citep{Adam06},
 which are also summarized in \citet{Itoh96,Sekiguchi10}.
  Hence $Q_{\nu,intr}^C$ can be expressed as,
\begin{eqnarray}
\label{eq:QintrC}
Q_{\nu,intr}^C&=&Q_{e^-}^{f}+Q_{e^-}^{h}+Q_{e^+}^{f}+Q_{e^+}^{h}\nonumber \\
&+&\sum_{\nu\in(\nu_e,\bar{\nu}_e,\nu_x)}
2(Q_{e^-e^+\rightarrow \nu\bar\nu}+Q_{\gamma\rightarrow \nu\bar\nu}+Q_{NN\rightarrow NN\nu\bar\nu}),
\end{eqnarray}
 where $Q_{e^{-/+}}^{f}$ and $Q_{e^{-/+}}^{h}$ represents the cooling rate
 by electron/positron capture on free nucleons and on heavy nuclei, respectively.

 Concerning the neutrino heating ($Q^{\mu,H}$), we only include 
 the dominant heating reactions in the gain region, which is 
 absorption of electron/anti-electron neutrinos by free nucleons.
 Then $Q^{\mu,H}$ reads
\begin{eqnarray}
\label{heat}
Q^{\mu,H}&=&e^{-\beta\tau}
\sum_{i\in(\nu_e,\bar{\nu}_e)}
\int d\omega\kappa_{\omega,i} \bigl(-\mathcal{J}_{\omega,i} u^\mu-
 \mathcal{H}_{\omega,i}^\mu\bigr)
\end{eqnarray}
where $\omega$ denotes neutrino energy,
$\kappa_{\omega,i}$ is the energy-dependent opacity for electron or 
 anti-electron neutrinos (i.e. $i = \nu_e$ or $\bar{\nu}_e$ see Appendix A),
  and $\mathcal{J_{\omega},H_{\omega}}$ is the energy-dependent
Eddington moments, respectively.
 Yielding to the prescription of the so-called light-bulb scheme (e.g.,
 \citet {Nordhaus10}),
 a term of $e^{-\beta\tau}$ is introduced to vanish
    the neutrino heating smoothly as the opacity becomes higher inward
  down to the neutrino sphere. 
 To take a gray approximation, we replace the energy integration in Equation
 (\ref{heat}) with the root-mean-squared (RMS) energy of the streaming neutrinos
  ($\epsilon_{s_{\nu},i}$, see Appendix A for
 the definition) as
\begin{eqnarray}
\label{eq:QmuH2}
 \int d\omega\kappa_{\omega,i} \bigl(-\mathcal{J}_{\omega,i} u^\mu-\mathcal{H}_{\omega,i}^\mu\bigr)\,\delta(\omega - \varepsilon_{s_\nu, i})
 \longrightarrow \ \ 
\bigr(\varepsilon_{s_\nu,i}\bigl)^2\tilde\kappa\bigl(-\mathcal{J} u^\mu-\mathcal{H}^\mu\bigr),
\end{eqnarray}
 where $\tilde\kappa$ denotes the monochromatic opacity, in which 
the energy-dependence is replaced with the one of the rms energy (namely, $\kappa_i 
= \tilde\kappa_i \cdot \varepsilon_{s_\nu,i}^2$)\footnote{The delta function in
 the left-hand-side of Equation (\ref{eq:QmuH2}) may be 
replaced by the Fermi-Dirac function. In the case, an additional factor of 
 $F_{2}(\eta_{\nu},0)$ (for the degeneracy limit $F_{2}(0,0) \approx 2$)
 can be multiplied, which could potentially enhance the impacts of neutrino heating.}.
  Since $\mathcal{J, H}$ in Equation (\ref{eq:QmuH2}) is related to the variables in the
  laboratory frame by Equations (\ref{j1},\ref{j2}), the two-moment equations of $E_{\nu}$ and
 $F_{\nu}$ with the source terms are finally closed.
 Having given explicit forms of $Q^{\mu}$,
   we are now moving on to summarize the GR hydrodynamic equations including 
 the source terms in the next section.
   
\subsection{Hydrodynamic Equations}
\label{sec:hydro}
From Equation (\ref{eq5}), the hydrodynamic
 equations are written in a conservative form as,
\begin{eqnarray}
\label{eq:GRmass}
\partial_t \rho_{\ast}+\partial_i(\rho_\ast v^i)&=&0,\\
\label{eq:GRmomentum}
\partial_t \hat S_i+\partial_j(\hat S_i v^j+\alpha e^{6\phi}P\delta_i^j)&=&
-\hat S_0\partial_i \alpha+\hat S_k\partial_i \beta^k+2\alpha e^{6\phi}S_k^k\partial_i \phi\nonumber \\
&&-\alpha e^{2\phi} ({S}_{jk}-P \gamma_{jk}) \partial_i 
\tilde{\gamma}^{jk}/2-e^{6\phi}\alpha Q^\mu \gamma_{i\mu},\\
\label{eq:GRenergy}
\partial_t \hat \tau+\partial_i (\hat S_0v^i+e^{6\phi}P(v^i+\beta^i)-\rho_\ast v^i)&=&
\alpha e^{6\phi} K S_k^k /3+\alpha e^{2\phi} ({S}_{ij}-P \gamma_{ij})\tilde{A^{ij}}-\hat S_iD^i\alpha\nonumber\\
&&+e^{6\phi}\alpha Q^\mu n_\mu,\\
\label{eq:GRlepton}
\partial_t (\rho_\ast Y_e)+\partial_i (\rho_\ast Y_e v^i)&=&\rho_\ast \Gamma_e,
\end{eqnarray}
where $\hat X\equiv e^{6\phi}X$, $\rho_\ast\equiv\rho We^{6\phi}$, $S_i\equiv\rho hW u_i $ and
$S_0\equiv\rho h W^2-p$.
$\rho$ is the rest mass density, $u_\mu$ is the 4-velocity of fluid, $h\equiv 1+\varepsilon+p/\rho$ is the specific enthalpy,
$v^i=u^i/u^t$, $\hat\tau=\hat S_0-\rho_{\ast}$, $Y_e$ is the electron fraction,
$\varepsilon$ and $p$ is the internal energy and pressure, respectively (see, Appendix \ref{sec:Implementation of EOS}).

From Equations (\ref{eq:QmuC}, \ref{eq:QmuH}),
the source terms appearing in the right-hand-side of Equations
 (\ref{eq:GRmomentum}, \ref{eq:GRenergy}) can
  be explicitly written as,
\begin{eqnarray}
\label{eq:QmuGamma_mui}
-Q^\mu\gamma_{\mu i}&=&-(Q^{\mu,C}-Q^{\mu,H})\gamma_{\mu i} \nonumber\\
&=&-\sum_{\nu\in\nu_e,\bar\nu_e,\nu_x}\Bigl [(1-e^{-\beta_\nu\tau_\nu})Q_{\nu,diff}+e^{-\beta_\nu\tau_\nu}Q_{\nu,intr}^C \Bigr]u_i\nonumber\\
&&+\sum_{\nu\in\nu_e,\bar\nu_e}e^{-\beta_\nu\tau_\nu}\bigr(\varepsilon_{s_\nu}\bigl)^2\tilde\kappa_\nu
\bigl(-WF_{\nu i}+P^k_{\nu i} u_k\bigr)\\
\label{eq:QmuNmu}
Q^\mu n_{\mu}&=&(Q^{\mu,C}-Q^{\mu,H})n_\mu \nonumber\\
&=&-\sum_{\nu\in\nu_e,\bar\nu_e,\nu_x}\Bigl [(1-e^{-\beta_\nu\tau_\nu})Q_{\nu,diff}+e^{-\beta_\nu\tau_\nu}Q_{\nu,intr}^C \Bigr]W\nonumber\\
&&+\sum_{\nu\in\nu_e,\bar\nu_e}e^{-\beta_\nu\tau_\nu}\bigr(\varepsilon_{s_\nu}\bigl)^2\tilde\kappa_\nu
\bigl(WE_\nu-F_\nu^k u_k\bigr).
\end{eqnarray}
 $\Gamma_e$ in Equation (\ref{eq:GRlepton}) denotes
    the change in $Y_e$ due to neutrino-matter
   interactions, which can be estimated in the same way as $Q^{\nu}$. 
 Given an appropriate EOS, the hydrodynamic equations
    (\ref{eq:GRmass})-(\ref{eq:GRenergy}) are closed. 
We employ the latest version of tabulated EOS by Shen et al.(98)\footnote{e.g., 
 Shen EOS (2011) downloadable from http://user.numazu-ct.ac.jp/~sumi/eos/} for 
heavy nuclei and uniform nuclear matter. Since Shen EOS contains contributions 
 only from baryons, we need to add contributions from electron/positron, 
 and photon (see Appendix B for more details).

\section{Initial Models and Numerical Methods}
\label{sec:Models and numerical methods}
\subsection{Initial Models}
To assess GR and 3D effects on the neutrino-heating mechanism,
we compute four models with a combination of SR or GR hydrodynamics
 in 1D or 3D, which we label as
 1D-SR, 1D-GR, 3D-SR, and 3D-GR, respectively.
 We employ a widely used progenitor of a 15$M_\odot$ star 
(\citet{WW95}, model ``s15s7b2'').
In our SR models, the space-time metric is assumed to be flat 
(i.e., $\alpha=1,\ \beta^i=0,\ \gamma_{ij}=\delta_{ij},\ \phi=0,\ K_{ij}=0$) and 
 we also assume that the self-gravity acts instantaneously in a Newtonian way 
which is evaluated by solving the following Poisson equation,
\begin{eqnarray}
\label{eq:SelfGravity}
\nabla^2\phi_{NT}=4\pi S_0.
\end{eqnarray}

We iteratively solve this huge simultaneous equation by the so-called ``BiConjugate Gradient Stabilized
(BiCGSTAB)" method \citep{vanderVorst92} with an appropriate boundary condition\footnote{The boundary condition is taken as $\phi_{NT}|_{\partial S}\equiv -M^0/r-M^ix_i/r^3-M^{ij}x_ix_j/r^5$
where $M^0$, $M^i$ and $M^{ij}$ are the monopole, dipole and quadrupole momenta of $S_0$, e.g.
$M^{ij}\equiv \int S_0 x^i x^j dV$.}.
Then the source term in SR associated with gravity appearing in the right-hand-side 
 of Equations (\ref{eq:GRmass})-(\ref{eq:GRenergy}) is explicitly written as
\begin{eqnarray}
\label{eq:SRsource_term}
{\bf S}=\{0,-\rho\partial_i\phi_{NT},-\rho v^i \partial_i\phi_{NT}\}.
\end{eqnarray}
In practice, we first evaluate $\rho v^i$ in Equation (\ref{eq:SRsource_term}) 
by the numerical flux of the rest mass density and $\partial_i\phi_{NT}$
defined at cell interfaces. We then take an average of the product, $\rho v^i \partial_i\phi_{NT}$, over the cell surfaces to evaluate the gravitational
source terms defined at the cell center (see \citet{Kuroda10} for further details 
such as about how to treat the self-gravity in the AMR structure and how to evolve the space-time metric with GR hydrodynamics).

To construct 1D models in our Cartesian code, the following condition 
for the spacial velocity $u_i$ (and also for $S_i$) is enforced at every (hydro-) 
timestep, 
\begin{eqnarray}
\label{eq:1DVr}
u_i=\frac{x^j u_j}{|x|^2}x^i.
\end{eqnarray}
As can be read, this operation eliminates the non-radial components of the 
 flow velocity and momentum.
 Although the artificial elimination 
 could potentially lead to the shift of the kinetic energy into the thermal one,
 our 1D results (without and with 
 neutrino heating/cooling) are in good agreement with
 the previous 1D results as will be mentioned in Appendix \ref{appC} and 
 section \ref{sec:Results}. This suggests that the manipulation
  is not severely bad for the sake of this study.
However, in our 1D-GR model, this artificial procedure violates the momentum 
constraint more or less, to which we will come back in Appendix \ref{appC}.

The 3D computational domain consists of a cube of 
 $10000^3$ km$^3$ volume in the Cartesian coordinates.
 In our 3D models, we set the maximum refinement AMR level ($L_{\rm AMR}$, i.e., refine
 the AMR boxes in the vicinity of the center) at 
 5 in the beginning and then increment it as the collapse proceeds.
 We define the criterion to increment $L_{\rm AMR}$ 
every time the central density exceeds $10^{12,13,13.5}$ g cm$^{-3}$ during the infall phase
(see \citet{Kuroda10} for more details). 
Each AMR level consists of $8^3$ AMR blocks with a nested structure and 
 each AMR block has $8^3$ cubic cells.
Roughly speaking, such structure corresponds to an angular resolution 
of $\sim$ 2 degrees through the entire computational volume.
 Near core bounce, an effective resolution becomes $\Delta x\sim 600$ m 
in the center of our 3D models. The numerical resolutions are summarized in Table \ref{tb:resolution}. We note that the employed resolutions in the central region are almost similar to those in 
 \citet{ott12} who very recently reported 3D GR results using AMR technique.
 However, our resolutions near the accretion shock surface ($r\sim100-150$km) are $
 \ga$ 2 times coarser than their value $\Delta x\sim900$m.

\begin{deluxetable}{cccccc} 
\tablecolumns{6} 
\tablewidth{0pc} 
\tablecaption{\label{tb:resolution}Numerical resolution of our 3D(-SR/GR) models
 near core bounce.
 The numerical resolution ($\Delta x$) is shown for different AMR levels (here from 
 the finest level of 8 down to 4). $r\la\sqrt3 |x|$ represents the box size 
corresponding to each AMR level.}
\tablehead{ 
\colhead{AMR level} & \colhead{8}   & \colhead{7}    & \colhead{6} & 
\colhead{5}    & \colhead{4}\\
\colhead{$\sqrt{3}|x|$} & \colhead{$\la 33.7$km}   & \colhead{$\la68$km}    & \colhead{$\la136$km} & 
\colhead{$\la272$km}    & \colhead{$\la544$km} }
\startdata 
$\Delta x$ & $\sim600$m & $\sim1.2$km & $\sim2.4$km & $\sim4.8$km & $\sim9.6$km  \\ 
\enddata 
\end{deluxetable}


\subsection{Numerical Methods}
  Since the hydrodynamic and transport equations  (Equations
  (\ref{eq:GRmass})-(\ref{eq:GRenergy}) and (\ref{rad1}, \ref{rad2}))
   are all expressed in a hyperbolic form, they can be evolved by a
    standard high-resolution-shock-capturing scheme. We utilize the  
HLL (Harten-Lax-van Leer) scheme
  \citep{Harten1983} to evaluate the numerical fluxes.
 A reconstruction of the primitive variables defined at immediate
 left/right of the cell surface is performed by a monotonized
 central method \citep{VanLeer77}.
The fastest (or right-going) and slowest (or left-going)
characteristic wave speeds of fluid system, $\lambda_{flu}$, for
$i(\in x,y,z)$ direction are obtained by solving the following second
 order equation,
 \begin{eqnarray}
\label{eq:LambdaFluid}
\mathcal{A}\lambda_{flu}^2+2\mathcal{B}\lambda_{flu}+\mathcal{C}=0
\end{eqnarray}
where
\begin{eqnarray}
\label{eq:LambdaFluidABC}
\mathcal{A}&=&\biggl(\frac{1}{c_s^2}-1\biggr)W^2+1, \nonumber \\
\mathcal{B}&=&\beta^i-\biggl(\frac{1}{c_s^2}-1\biggr)(\alpha v^i-\beta^i)W^2, \nonumber \\
\mathcal{C}&=&\biggl(\frac{1}{c_s^2}-1\biggr)W^2(\alpha v^i-\beta^i)^2+{\beta^i}^2-\alpha^2\gamma^{ii},
\end{eqnarray}
and $c_s$ is the sound velocity (see Appendix B.3).

Meanwhile, the fastest and slowest characteristic wave speeds of radiation system, 
$\lambda_{rad}$, are assumed to have the same expression of the radiation pressure 
 (Equation (\ref{neu_p})) as
\begin{eqnarray}
\label{eq:LambdaRad}
\lambda_{\rm rad}=\frac{3\chi-1}{2}\lambda_{\rm rad,thin}+\frac{3(1-\chi)}{2}
\lambda_{\rm rad,thick},
\end{eqnarray}
where $\lambda_{\rm rad,thin}$ and $\lambda_{\rm rad,thick}$ is determined 
by $P^{ij}_{\rm thin}$ and $P^{ij}_{\rm thick}$, respectively.
 According to \cite{Shibata11}, the fastest (slowest) wave speed in the 
 optically thick or thin limit is evaluated by taking maximum (minimum) values, that is,
\begin{eqnarray}
\label{eq:LambdaThick}
\Biggl(-\beta^i +\frac{2W^2p^i\pm\sqrt{\alpha^2\gamma^{ii}(2W^2+1)-2W^2{p^i}^2}}{2W^2+1},-\beta^i+ p^i\Biggr),
\end{eqnarray}
for the optically thick limit (where $p^i=\gamma^{ij}u_j/u^t$) and
\begin{eqnarray}
\label{eq:LambdaThin}
\Biggl(-\beta^i\pm \alpha\frac{F^i}{\sqrt{F_jF^j}},-\beta^i+ \alpha E\frac{F^i}{F_jF^j}\Biggr),
\end{eqnarray}
for the optically thin limit, respectively.
 With these wave velocities regarding the fluid and radiation component
($\lambda_{flu/rad}$), we define the HLL flux \citep{Anton06} as
 \begin{eqnarray}
\label{eq:FHLL}
{\bf F}_{HLL}=\frac{\tilde\lambda_+{\bf F}_L-\tilde\lambda_-{\bf F}_R+\tilde\lambda_-\tilde\lambda_+({\bf Q}_R-{\bf Q}_L)}
{\tilde\lambda_+-\tilde\lambda_-},
\end{eqnarray}
where $\tilde\lambda=\lambda/\alpha$, 
$L/R$ denotes the left/right states for the Riemann problem with 
 ${\bf F}_{L/R}$ and ${\bf Q}_{L/R}$ representing the advection and conservative terms,
 respectively.

To ensure conservative laws at the interface of different AMR levels,
we furthermore need to perform a ``{\it refluxing}'' procedure in estimating 
the numerical flux \citep[see][]{Kuroda10}.
To evolve the BSSN terms, we adopt the 4th order finite differencing for the spatial
derivatives and the 4th order upwind differencing for the advection terms
\citep{Zlochower05,Etienne08} except at the AMR boundary.
At the AMR boundary, we employed 3rd order upwind scheme for the advection terms.
 Numerical tests are presented in Appendix 
\ref{appC}, in which we first show a 1D adiabatic core-collapse test 
to validate the implementation of Shen EOS in the code, followed by 
 the corresponding 1D tests including neutrinos.

\section{Results}
First let us compare prebounce features among the four models 
(1D-SR, 1D-GR, 3D-SR, and 3D-GR) in section \ref{4_1}, and move on 
 to the postbounce phase in section \ref{4_2}.
 Then in section \ref{4_3}, we will discuss the 3D/GR effects on 
 the neutrino-heating mechanism.

\label{sec:Results}
\subsection{Infall and Bounce}
\label{4_1}

\begin{figure}[htpb]
\begin{center}
\includegraphics[width=60mm,angle=-90.]{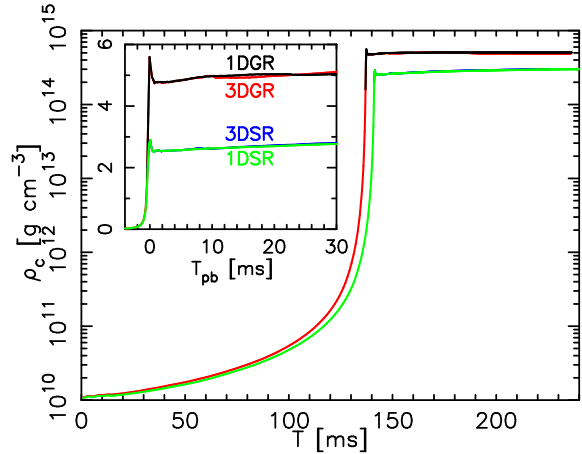}
\end{center}
  \caption{Central density $\rho_c$ as a function of 
 time from initial collapse for our models of 3D-GR ({\it red}), 
1D-GR ({\it black}), 3D-SR ({\it blue}), and 1D-SR ({\it green}), respectively.
  The inset is just zoom up near bounce, in which the time is measured from bounce 
($t_{\rm pb} \equiv$ 0) and  the vertical lines represent
 the central density normalized by $10^{14}$ g cm$^{-3}$ in a linear scale.}
\label{pic:F1}
\end{figure}

\placefigure{pic:F1}

We begin our comparisons with the infall, bounce, and immediate postbounce phase.
As seen from Figure \ref{pic:F1}, collapse to bounce takes slightly less time 
 in our GR models (137 ms) compared to the SR models (141 ms),
 and the central density $\rho_c$ at bounce is approximately 2 times larger in the 
GR simulations than in the corresponding SR simulations (see the inset in Figure 
 \ref{pic:F1}).
 
  For our non-rotating progenitor, the dynamics of collapsing iron core 
 proceeds totally spherically till the stall of the bounce shock. This is the reason
 that the multi-D effects are invisible in the immediate 
 postbounce phase. Hence
 we focus on the comparisons between the 1D-SR and 1D-GR model in the rest of this
 subsection.

\begin{figure}[htpb]
\begin{center}
\includegraphics[width=60mm,angle=-90.]{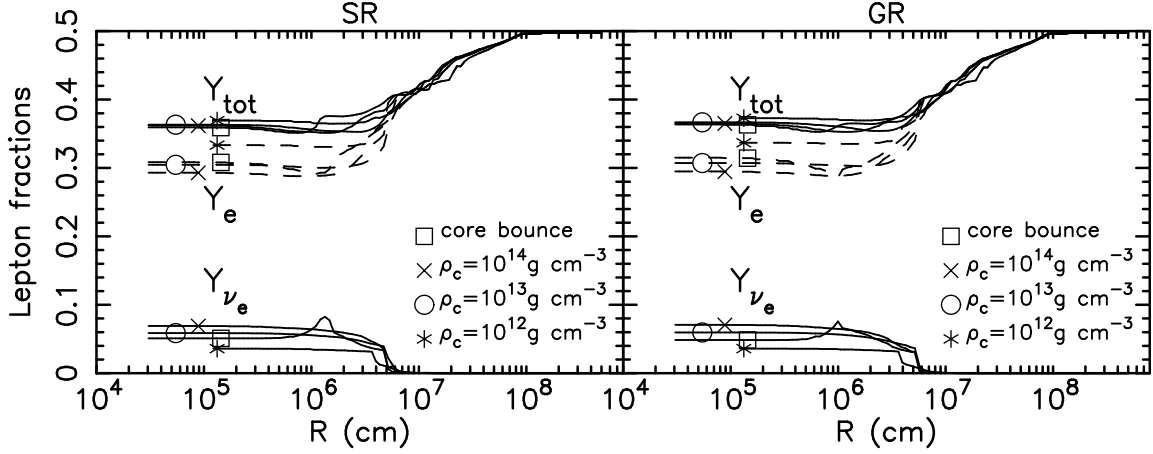}
\end{center}
  \caption{Profiles of the total lepton fraction ($Y_{\rm total}$),
 electron fraction ($Y_e$), and electron-type neutrino fraction ($Y_{\nu_e}$) for 
 the 1D-SR (left panel) and 1D-GR model (right panel) at times, when the central 
 density reaches the value as indicated in the plots.
 Profiles of the electron fraction are plotted by {\it dashed} lines to distinguish from those of total lepton
 fraction.}
\label{f2}
\end{figure}

Figure \ref{f2} shows several snapshots of the lepton fraction ($Y_{\rm total}$),
 electron fraction ($Y_e$), and electron-type neutrino fraction ($Y_{\nu_e}$) for
 the 1D-SR (left panel) and 1D-GR model (right panel), respectively.
After neutrino trapping (i.e. at a central density of a few $10^{12}$ g cm$^{-3}$), 
 the central lepton fraction (black lines) is shown to be conserved 
later on. In the trapped regions, the radial profile of the neutrino fraction 
(blue lines) is almost flat, while $Y_{\nu_e}$ shows a gradual increase to satisfy the $\beta-$equilibrium.

\begin{figure}[htpb]
\begin{center}
\includegraphics[width=60mm,angle=-90.]{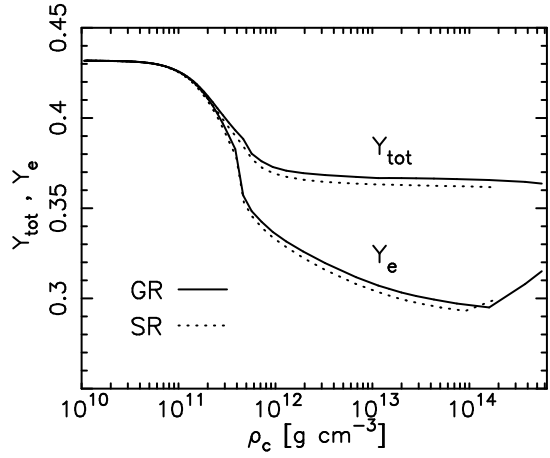}
\end{center}
  \caption{Comparison of the electron and lepton fraction versus
 central density during collapse for the 1D-SR (solid line) and 1D-GR (dotted line)
 model, respectively.}
\label{f3}
\end{figure}

 From Figure \ref{f3}, it can be seen that the lepton fraction at bounce (right 
end-point in density)
 is slightly larger for the 1D-GR model ($\sim0.364$, solid line) 
compared to the 1D-SR model ($\sim0.359$,
 dotted line). The slight suppression of deleptonization is
 possibly because the neutrino opacity is effectively enhanced because 
 of the more compact core in the GR model.
 Note that this trend is qualitatively in accord with the previous 1D results in which 
 a spectral neutrino transport was solved (e.g., \citet{Lentz11,Bernhard10}). Quantitatively, the central density ($\rho_c = 5.5\times10^{14}$ g cm$^{-3}$) 
at bounce in our 1D-GR model is higher than the one ($\rho_c = 3.4\times10^{14}$ g cm$^{-3}$) obtained in a 1D full-fledged GR
 simulation by \cite{Sumiyoshi05} who solved the 1D Boltzmann transport using the same 
progenitor and the same EOS. Regardless of big differences in the transport 
 scheme as well as employed microphysics (i.e., very detailed vs. very approximate), 
the central $Y_e \approx 0.3$ at bounce happens to be very close with each other. 
 Our results on the 1D-GR model are rather similar to \cite{Sekiguchi10}. This is 
 reasonable in the sense that the our neutrino transport relies partly on 
 a multi-species leakage scheme using similar microphysical detail.

\placefigure{f2}
\placefigure{f3}

\begin{figure}[htpb]
\begin{center}
\includegraphics[width=60mm,angle=-90.]{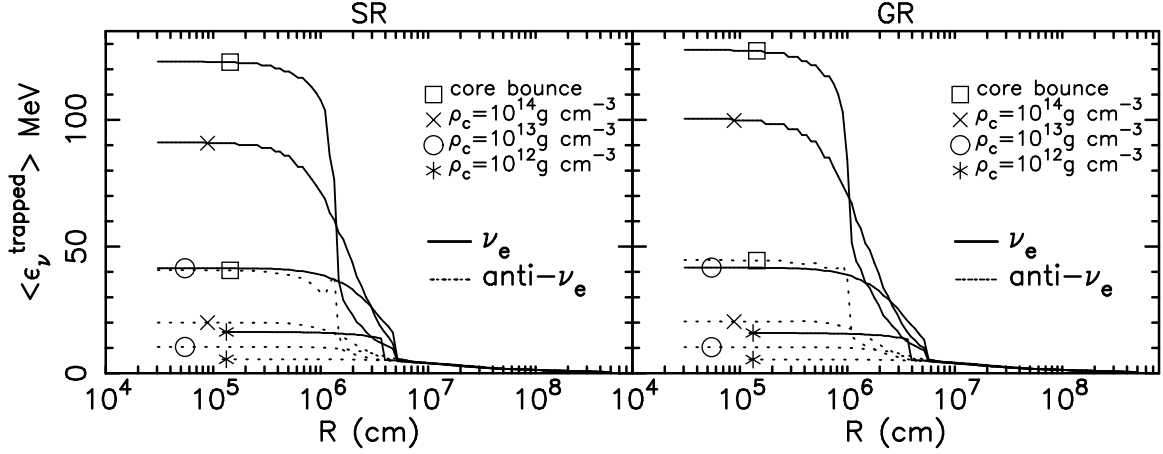}
\end{center}
  \caption{Same as Figure \ref{f2}, but for the profiles of the trapped neutrino 
energies for $\nu_e$ ({\it solid}) and $\bar\nu_e$ ({\it dotted}).}
\label{pic:AppF5_E}
\end{figure}
\begin{figure}[htpb]
\begin{center}
\includegraphics[width=70mm,angle=-90.]{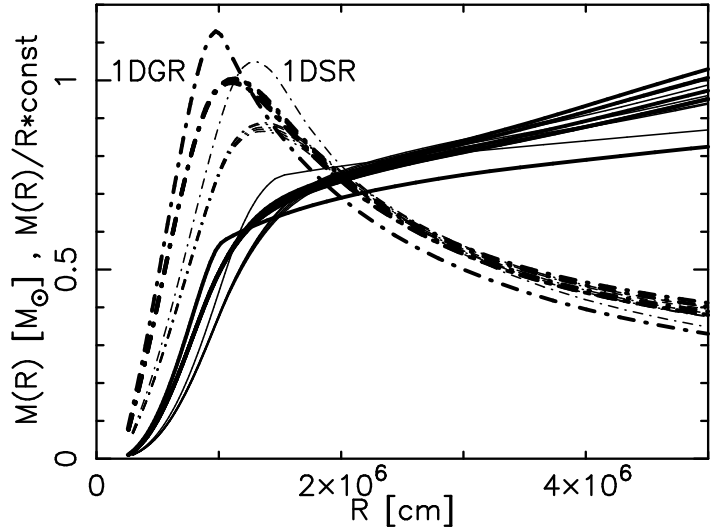}
\end{center}
  \caption{Profiles of enclosed mass $M(R)$ ({\it solid}) and 
the compactness parameter $M(R)/R$ ({\it dash}-{\it dotted}) as a function of radius $R$,
 in which lines are drawn every 2 ms in the first 10 ms postbounce for the 1D-SR 
 (thin line) and 1D-GR (thick line) model, respectively.
 The 1D-GR model has the maximum compactness parameter 
 that is 10\% larger, and the mass of the homologous core at bounce 
  that is 20 \% smaller compared to the 1D-SR model.}
\label{pic:F6_sub}
\end{figure}

 Figure \ref{pic:AppF5_E} compares the mean energy of trapped neutrinos 
$\langle\varepsilon_\nu \rangle$ between the 1D-SR (left) and 1D-GR model (right)
 (see Appendix A for definition).
 The mean energy is shown to be 20\% larger (maximally near at bounce) for
 the GR model compared to the SR counterpart.
This is also because of the more compact core due to the GR 
 hydrodynamics (Figure \ref{pic:F6_sub}), 
leading to a more hotter neutrino sphere at smaller radii.

\placefigure{pic:AppF5_E}
\placefigure{pic:F6_sub}


\clearpage

\subsection{3D hydrodynamics in the postbounce phase}
\label{4_2}

\begin{figure}[htpb]
\begin{center}
\includegraphics[width=80mm,angle=-90.]{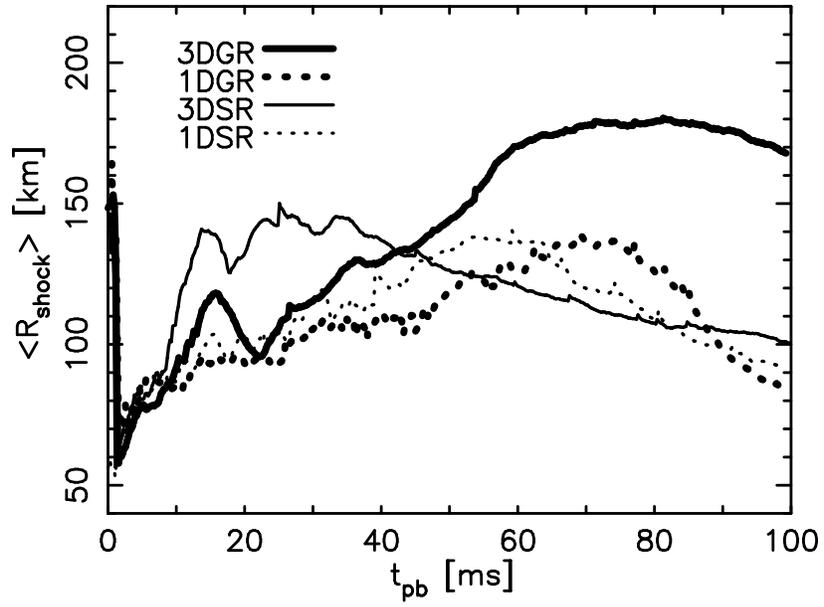}
\end{center}
  \caption{Evolutions of average shock radii as a function of post-bounce time 
$t_{\rm pb}$ for the four variant models.}
\label{shock}
\end{figure}

\begin{figure}[htbp]
\begin{center}
\includegraphics[width=150mm]{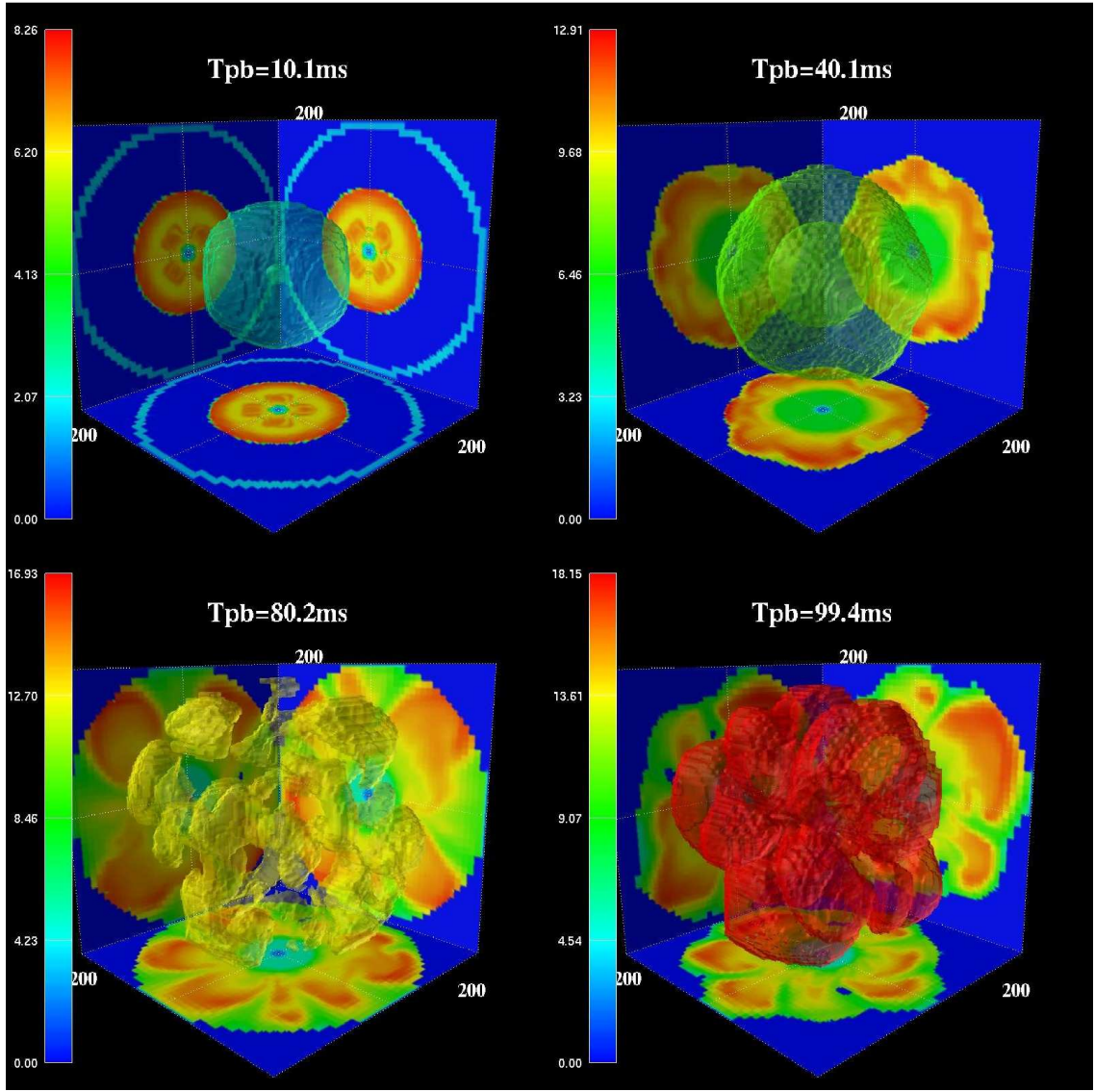}
\end{center}
\caption{Three dimensional plots of entropy per baryon for four snapshots 
(top left; $t_{\rm pb} = 10$ ms, top right; $t_{\rm pb} =40$ ms, bottom left; $t_{\rm pb} =80$ ms,
 and bottom right; $t_{\rm pb} =100$ ms)
 for the 3D-GR model.
The contours on the cross sections in the
$x=0$ (back right), $y=0$ (back bottom), and $z=0$ (back left) planes are, 
respectively projected on the sidewalls of the graphs to visualize 
3D structures. For each snapshot, 
 the arbitrary chosen iso-entropy surface is shown, and 
the linear scale is indicated along the axis in unit of km. }
\label{f7}
\end{figure}

\placefigure{shock}
\placefigure{f7}


In the first 10 ms after bounce, the bounce shock turns into the so-called ``passive'' 
shock, which expands 
gradually outward with no positive radial velocities \citep{Buras06a}. 
 As shown in Figure \ref{shock}, the average shock radii until the passive 
 expansion starts ($t_{\rm pb} \lesssim 10$ ms),
 remain almost close in all the models. It then diverges,
 which is more remarkable between the 3D models (solid lines). 
As will be discussed in the following, this is because
  multi-D effects (convection and SASI) sensitively affect the postshock 
hydrodynamic behaviors, also under the influence of the different treatment 
 in gravity (SR versus GR).

First of all, let us compare the shock evolution among the 1D models in SR vs.
 GR (1D-SR (thin dotted line) or 1D-GR (thick dotted line)).
As expected, the shock radius 
is generally more compact for the 1D-GR model (thick dotted line). 
On the other hand, 
the maximum shock extent is observed to be almost the same with each other 
($\langle R_{\rm shock} 
 \rangle \sim 130$ km). Though we cannot unambiguously specify the reason, 
this trend was also seen in \citet{Bernhard10}, who 
compared the shock radii in 1D 
simulations with detailed neutrino transport 
 in CFC vs.
 the corresponding Newtonian model with the effective potential approach 
(e.g., right panel of their Figure 5).
 The maximum of $\langle R_{\rm shock} \rangle$ in Figure 
\ref{shock} for our 1D models 
 indicates the epoch when the passive expansion stops. Afterwards
 ($t_{\rm pb} \gtrsim 70$ ms), the shock begins to shrink and  
a much more rapid recession is visible for the GR model (thick dotted line).
 The maximum shock radii and the shock recession timescale obtained here
are again similar to those obtained in the previous 1D results 
for the same progenitor model with employing the Shen EOS \citep{Sumiyoshi05}.
 The maximum shock extent was shown in \citet{Sumiyoshi05} at 
  a radius of $\sim 150$ km at around 80-90 ms postbounce, which is at $\sim 140$ km
 at $\sim 70$ ms postbounce in our 1D-GR model. Considering the mentioned difference 
 of neutrino transport as well as different hydrodynamics solvers, it may not be 
 so unreasonable to say that our results show a relatively good agreement 
with \cite{Sumiyoshi05}. In addition, \cite{Sumiyoshi05} showed that the
 shock propagation during the first $\sim200$ ms after bounce does not depend so much 
on the EOS. This should be the reason that our 1D-GR results, at least for 
 the evolution of the shock radii, are quite similar to \citet{Lentz11,Bernhard10}.

\begin{figure}[htbp]
\begin{center}
\includegraphics[width=90mm,angle=-90.]{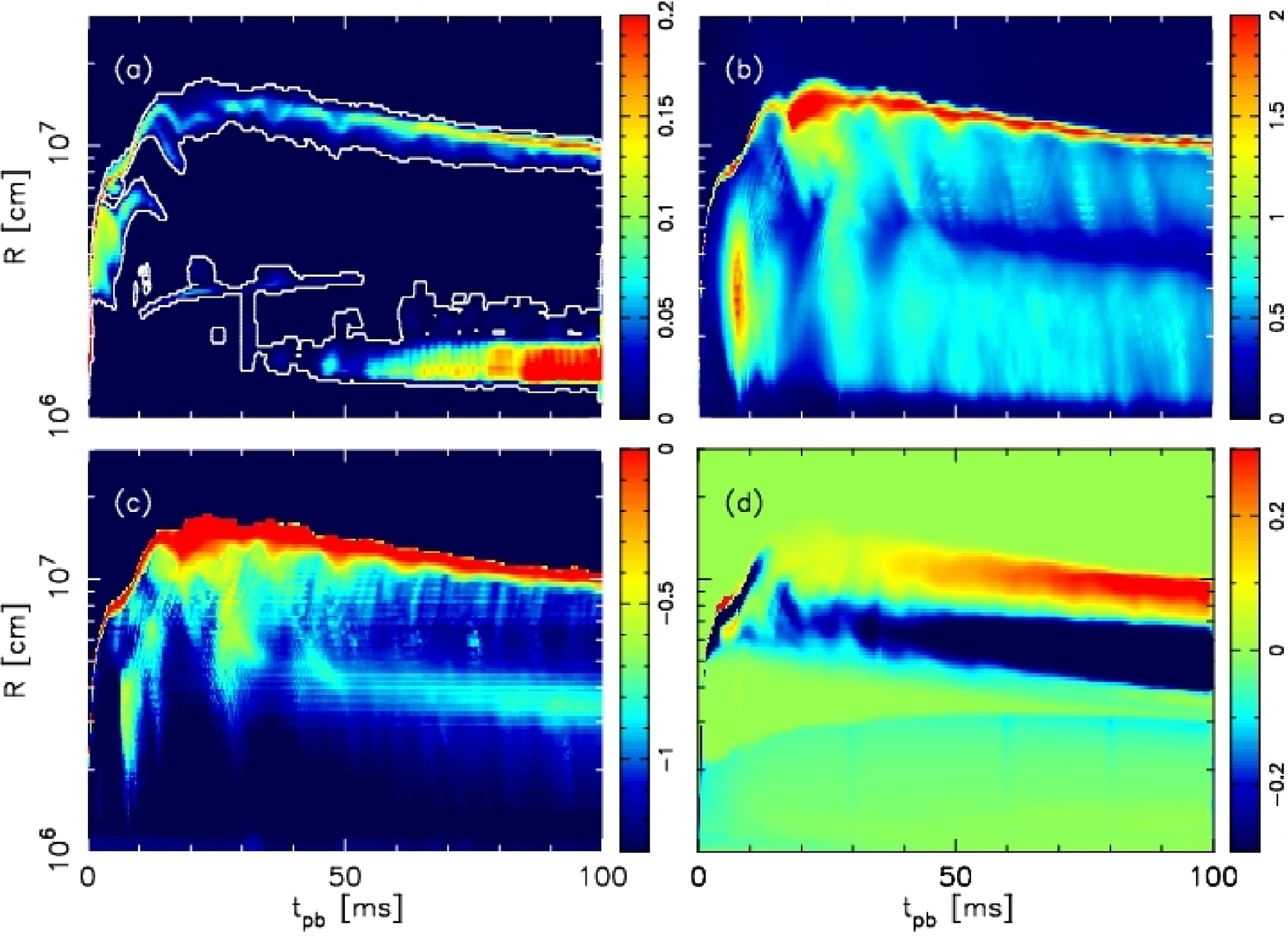}\\
\includegraphics[width=125mm]{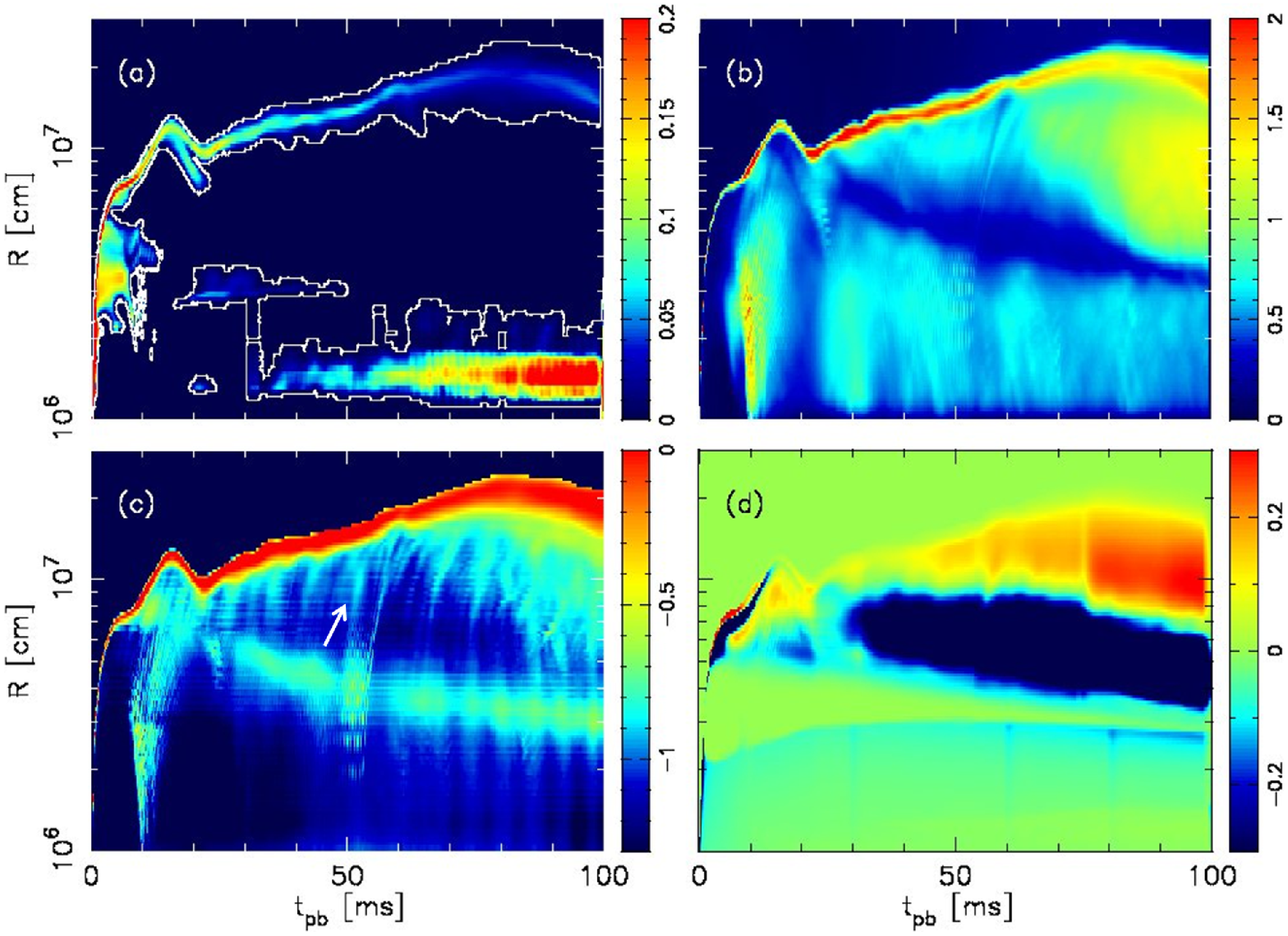}
\end{center}
\caption{Panels (a) to (d) display angle averaged; (a) Brunt-V$\ddot{\rm a}$is$\ddot{\rm a}$l$\ddot{\rm a}$ frequency ($\omega_{\rm BV}$ ms$^{-1}$); 
(b) anisotropic velocity $V_{\rm aniso}$ normalized by $10^9$ cm s$^{-1}$; (c) normalized pressure perturbation $\Delta p$ 
(in a logarithmic scale) (d) net energy deposition rate per baryon Q$_{\rm net}$ [GeV nuc$^{-1}$ s$^{-1}$] for our 3D-SR (top four panels)
 and 3D-GR model (the rest four), respectively. Note that convectively unstable regions (i.e., $\omega_{\rm BV}>0$) are only shown in panel (a)
  and the white line represents the contour of $\omega_{\rm BV}=0$.
  In panel (d), color contour of negative value of Q$_{\rm net}$ is saturated at -0.3.
 To guide the eye, a white arrow is inserted in
 panel (c) which points to a up-going pressure perturbation to the shock.}
  \label{pic:F7}
\end{figure}

\placefigure{pic:F7}

Four snapshots in Figure \ref{f7} are helpful to characterize 
 the postbounce features in our 3D-GR model. The top left 
panel is for $t_{\rm pb} \approx 10$ ms, when the bounce shock stalls 
at a radius of $\sim 90$ km (seen as a central blueish sphere). From the sidewall panels, 
the dominance of the $\ell = 4$ and $m=4$ mode can be seen in the postshock region,
 which is a numerical artifact inherent to the use of Cartesian coordinates. 
 Comparing the top left to top right panel in Figure \ref{f7}, 
the size of the outer sphere 
that marks the position of the shock (seen as greenish 
in the top right panel) becomes bigger, which is due to 
 the passive expansion. At this stage, there forms the gain region in which 
 neutrino heating dominates over neutrino cooling (e.g., \citet{Janka01}).
 The neutrino-driven convection gradually develops later on. The sidewall panels of the top right panel also indicate
 the growth of the postshock convection triggered by Rayleigh-Taylor instabilities.
 The entropy behind the standing shock becomes 
higher with time due to neutrino-heating, which can be inferred from
 a yellowish bubble in the bottom left panel.
 The high entropy bubbles ($s[k_B/{\rm baryon}]\ga10$) rise and sink behind the standing shock. The shock deformation is dominated by unipolar and bipolar modes, 
which may be interpreted as an onset of the SASI.
  The size of the neutrino-heated regions 
grows bigger with time in a non-axisymmetric way, which is indicated by  
 bubbly structures with increasing entropy (indicated by reddish regions
 in the bottom right panel).

During our simulation time (100 ms after bounce), the shock radii can 
 reach most further out for our 3D-GR model (red line in Figure \ref{shock}).
 In contrast, the shock has already shown a trend of recession in other models.
  Before we focus on the reason of it in the final section, 
let us next compare the activities of convective overturns as well as possible
 onset of the SASI that we only touched on above.

Figure \ref{pic:F7} displays space-time diagrams of laterally averaged; 
(a) Brunt-V$\ddot{\rm a}$is$\ddot{\rm a}$l$\ddot{\rm a}$ (BV) 
frequency $\omega_{\rm BV}$ (top left panel) ; (b) the anisotropic velocity 
$V_{\rm aniso}$(top right);  (c) the pressure perturbation $\Delta p$ in 
 a logarithmic scale (bottom left), and (d) the net heating rate per baryon 
Q$_{\rm net}$ (bottom right) for our 3D-SR (top four panels) and 3D-GR models 
(the other four panels), respectively.
Each of the quantities is defined as,
\begin{eqnarray}
\label{eq:OmegaBV}
\omega_{\rm BV}\equiv {\rm sign}(C_{\rm L})\sqrt{|g_{\rm eff}C_{\rm L}|},
\end{eqnarray}
 where {\bf $g_{\rm eff}$ represents effective gravitational acceleration that is estimated
 by taking a radial gradient of the potential, i.e. $g_{\rm eff}=d\phi_{NT}/dr$ for SR models
 and $g_{\rm eff}=d\alpha/dr$ for GR models, respectively}.
$C_{\rm L}$ is the Ledoux criterion;
\begin{eqnarray}
\label{eq:LedouxCriterion}
C_{\rm L}\equiv -\frac{\partial\rho}{\partial P}\biggl|_{s,Y_{\rm tot}}\biggl(\frac{\partial P}{\partial s}\biggl|_{\rho,Y_{\rm tot}}\frac{ds}{dr}+\frac{\partial P}{\partial Y_{\rm tot}}\biggl|_{\rho,s}\frac{dY_{\rm tot}}{dr}\biggr),
\end{eqnarray}
in which the neutrino contribution to entropy is taken into account where the 
$\beta$-equilibrium is satisfied \citep{Buras06a}.
 Following \citet{Takiwaki11}, $V_{\rm aniso}$ is estimated as
\begin{eqnarray}
\label{eq:Vaniso}
V_{\rm aniso}=\sqrt{\langle\rho\bigl[(v_r-\langle v_r\rangle)^2+v_\theta^2+v_\phi^2\bigr]\rangle/\langle\rho\rangle},
\end{eqnarray}
where  $\langle A \rangle $ represents the angle average of quantity $A$.
 We define the normalized pressure perturbation $\Delta p$ and the net heating rate per nucleon Q$_{\rm net}$ as
\begin{eqnarray}
\label{eq:DeltaP}
\Delta p\equiv \frac{\sqrt{\langle p^2 \rangle -\langle p \rangle^2}}{\langle p \rangle},
\end{eqnarray}
and
\begin{eqnarray}
\label{eq:QnetPerBaryon}
{\rm Q_{net}}\equiv \frac{e^{6\phi}\alpha Q^\mu n_\mu}{\rho},
\end{eqnarray}
respectively. At first glance of Figure \ref{pic:F7}, one may not see any big 
differences between 
the 3D-SR (top four panels) and 3D GR models (the other), but indeed there are. 
Let us first discuss the properties of the four panels (a) - (d) taking 
 the 3D-SR model as a reference and then proceed to focus on
 the differences between SR and GR. 

From panel (a) (top left) showing the BV frequency for 
the 3D-SR model, one can depict three typical convectively unstable regions 
in the postbounce phase;
   prompt convection (greenish region at $t_{\rm pb} \lesssim 20$ ms 
behind the shock\footnote{Note that the shock is indicated by 
 a white thin line quickly rising after bounce and the passive shock stalls at a radius of $R \sim 150$km
.},  postshock convection (seen as a narrow horizontal stripe behind the shrinking
 shock (just behind the outer most boundary labeled by white line),
 and PNS convection (clearly seen as a thick horizontal stripe above the 
 PNS at a radius of $\sim 10 - 20$ km later than $t_{\rm pb} \gtrsim 60$ ms).

\begin{figure}[htpb]
\begin{center}
\includegraphics[width=50mm,angle=-90.]{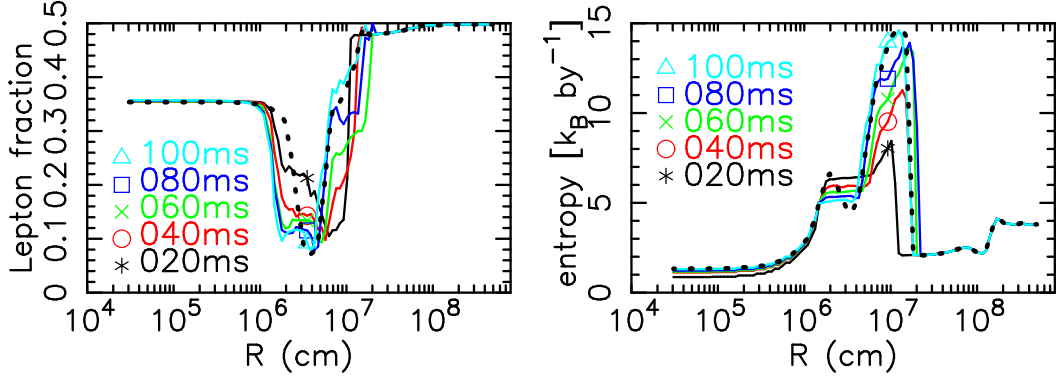}
\end{center}
  \caption{Profiles of angle-averaged total lepton fraction ({\it left}) and 
entropy ({\it right}) at times for the 3D-GR model, when the postbounce time 
is as indicated in the plots. Note that 
 the result of 1D-GR model at $t_{\rm pb}=60$ ms (black dotted curve)
 is shown for comparison.}
\label{pic:F10}
\end{figure}

In our 3D results, the PNS convection develops only very weakly
 before $\sim 60$ ms postbounce. This is due to the stabilizing effect 
by a positive entropy gradient (see the positive gradient persisting outside
 the PNS surface ($R \sim 10$ km) in the right panel of Figure \ref{pic:F10}). 
Afterwards, the PNS convection gradually becomes vigorous with 
 time as the negative lepton gradient nascent the PNS becomes remarkable 
 (see the steepening slope of $Y_l$ near $R \sim 10-20$ km in the left panel of 
Figure \ref{pic:F10}). Comparing the black dotted line in Figure \ref{pic:F10} 
(for 1D-GR model at $t_{\rm pb} = 60$ ms) with the corresponding one (green line) 
for the 3D counterpart, the slope of the negative gradient
 is shown to become much smaller for the 3D model both in the profiles 
 of lepton fraction (left panel) and entropy (right panel). 
This is a natural outcome of the convective overturns acting  
to wash out the local gradients.

From panel (b) in Figure 
\ref{pic:F10} showing the anisotropic velocity, 
the postshock convection ($R\ga100$km) is clearly seen as a reddish 
stripe running from top left to bottom right. 
Note in the panel that the prompt convection can be 
also seen like a narrow prolate spheroid colored by red 
at $t_{\rm pb} \lesssim 20$ ms with $10 \lesssim R \lesssim 60$ km. As seen,
 convective overturns operate above the PNS ($\sim 10-20$ km in radius)
 and below the shock $\sim 100$ km in radius. In-between, the region with 
 smaller anisotropic velocity is formed (seen as a horizontal stripe colored 
by deep-blue at a radius of $30 - 50$ km after $t_{\rm pb} \sim 50$ ms).
 By comparing to panel (d) (the net heating rate) to panel (c), the 
 region is overlapped with the cooling region (Q$_{\rm net} < 0$).
 The smaller anisotropic
 velocity there is because the infalling velocities in the cooling layer
are so high that the convectively unstable material cannot stay there for long.
 Such a configuration has been already presented in 2D \citep{Buras06a} 
and 3D results \citep{Takiwaki11}.

\placefigure{pic:F10}

Here let's see panel (c) not for the 3D-SR model but 
 for the 3D-GR model for convenience.
 The accreting flows should receive an abrupt deceleration near at the bottom of the cooling layer (the dark colored region in panel (d) (bottom right)), 
below which the regions are convectively stable (panel (b)).
 There forms a strong pressure perturbation (seen as a greenish horizontal stripe 
near $R\sim 30$ km in panel (c)). Subsequently the pressure perturbations propagate
 outward before they hit the shock (panel (c)), maybe leading 
 to the formation of the next vortices.
 These features seem at least not to be inconsistent with the
so-called advectic-acoustic cycle (e.g., \citet{Foglizzo00,Foglizzo02,Scheck08}
 and references therein), which is also observed in our 3D-SR model (top
 four panels).

\begin{figure}[htbp]
\begin{center}
\includegraphics[width=70mm,angle=-90.]{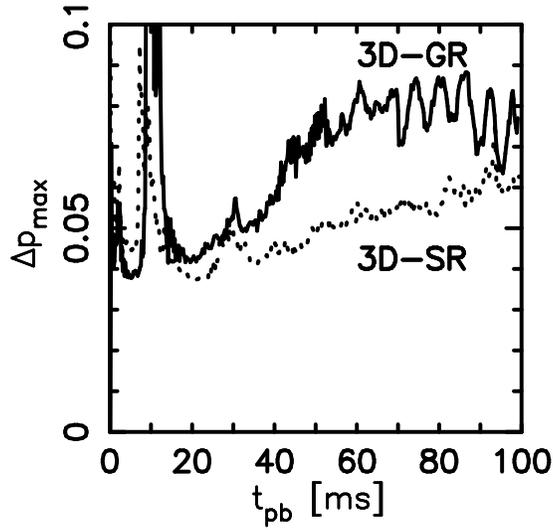}
\end{center}
\caption{Evolution of the maximum of the pressure perturbation
 $\Delta p_{\rm max}$ in our 3D-SR and 3D-GR models. 
In taking the maximum, we set the radial range 
 as $20\le R\le50$ km to cover the coupling radius 
(see panels (c) in Figure \ref{pic:F7}).}
\label{pic:F14}
\end{figure}

We now focus on the differences between the 3D-SR and 3D-GR models.
 Comparing the panel (a)'s between SR and GR, the unshocked core (regions below the PNS 
 convection at $t_{\rm pb} \gtrsim 50$ ms) is shown to be 
 more compact for the GR model. Between the pair models, 
 Figure \ref{pic:F14} compares 
 the maximum of the pressure perturbation that the 
 advecting vortices form near in the vicinity of the deceleration regions\footnote{\citet{Scheck08} 
 termed it as the ``coupling radius" in which the coupling of vortices and acoustic 
waves occur.}.  As seen, the pressure perturbation after bounce is generally
 larger for the GR model (solid line) compared to the SR model (dotted line) 
in our simulation time. This is presumably because stronger gravitation pull in GR 
makes the position of the coupling radius deeper, leading to produce
 more energetic acoustic waves. It is not straightforward to say something very 
 solid only from the figure, but what we observed in our 3D-GR model (i.e. generation of 
  stronger acoustic waves and the largest shock extent compared 
 to the SR counterpart) does not seem, at least, unfavorable to drive
 neutrino-driven explosions. In the next section, we move on to discuss more in detail
 how 3D and GR would potentially impact on the neutrino-heating mechanism.

\subsection{3D versus GR effects on the neutrino-heating mechanism}
\label{4_3}

Recalling that the neutrino heating rate can be 
symbolically expressed as $Q^{+}_{\nu} \propto L_{\nu}\langle \epsilon_{\nu}^2 \rangle$
 (e.g., \citet{Janka01}), we first
analyze the neutrino luminosities ($L_{\nu}$) and the mean 
energies ($\langle\varepsilon_\nu\rangle$) in the following.
 After that, we compare the dwell time to the neutrino-heating time in the gain region and
discuss which one (3D-SR vs. 3D-GR) is most likely to satisfy 
the criterion to initial the neutrino-driven explosions.

Figure \ref{lnu} shows evolution of the neutrino
 luminosities of all the species (for $\nu_e$, $\nu_x$ (left panel), and
 ${\bar{\nu}}_e$ (right panel)) for all the computed models. 
Here the neutrino luminosity is calculated as
\begin{eqnarray}
\label{eq:Lnu}
L_\nu\equiv\int{\alpha e^{6\phi}Q^{\mu,C}n_\mu dx^3},
\end{eqnarray}
where $Q^{\mu,C}$ in Equation (\ref{eq:QmuC}) takes into account all the cooling 
 contributions.

\placefigure{lnu}

The spike in the $\nu_e$ luminosity corresponds to 
the so-called neutronization, when the shock propagates out through the
 $\nu_e$ sphere.
 The peak $\nu_e$ luminosity for the GR models is 
$L_{\nu_e}\sim3\times10^{53}$ erg s$^{-1}$ (insensitve to 1D or 3D), which 
 is slightly luminous compared to those
 in the SR models ($L_{\nu_e}\sim2.9\times10^{53}$ 
erg s$^{-1}$). Using the same progenitor \citep{WW95}, this trend is 
qualitatively similar to \citet{Bruenn01}. On the other hand,  
  recent studies in which more detailed weak interactions are included
 in the Boltzmann transport have shown that the peak $\nu_e$ 
luminosity becomes $\sim 10$\% smaller for the GR models
 (e.g., \citet{Lentz11,Bernhard10}). This may carry an important message
  that the Boltzmann transport should be implemented in the full GR 
 simulations to obtain a $\sim 10$\%-order accuracy, which is not small
 at all when speaking about the neutrino-driven mechanism.
  
After the neutronization burst ($t_{\rm pb} \sim 10$ ms), the $\nu_e$ luminosity 
for the GR models slightly increases 
later on, while it stays almost constant
 for the SR models during the simulation time (green and blue lines). The $\bar{\nu}_e$
 luminosity after $50$ ms postbounce (right panel in Figure \ref{lnu})
 is highest for the 3D-GR model (red line),
 which is also the case for the $\nu_x$ luminosity (left panel).
 Although the luminosities change with time, the luminosities generally yield
 to the following order, 
\begin{itemize}
\item[]{for $\nu_e$, 3D-GR $>$ 1DGR, 3D-SR $\sim$ 1D-SR,}
\item[]{for $\bar{\nu}_e$, 3D-GR $>$ 1DGR, 3D-SR $>$ 1D-SR,}
\item[]{for ${\nu}_x$, 3D-GR $>$ 1DGR, 3D-SR $>$ 1D-SR.}
\end{itemize}
 To summarize, both 3D and GR work to raise the neutrino luminosities
 in the early postbounce phase. As seen from the left panel in Figure \ref{lnu}, 
GR maximally increases the $\nu_x$ luminosity up to $\sim 50 \%$ (in 3D),
 while the maximum increase by 3D is less than $\sim 20 \%$ (compare the $\bar{\nu}_e$ 
luminosity between the 3D-GR and 1D-GR model). These results indicate
 that compared to the spacial dimensionality, GR holds the key importance to enhance 
 the neutrino luminosities.

By comparing our 1D-GR results with those in \cite{Sumiyoshi05} again,
 the peak $\nu_e$ luminosity obtained here ($\sim 3.0\times10^{53}$ ergs s$^{-1}$)
 is higher than their Boltzmann results ($\sim1.8\times10^{53}$ ergs s$^{-1}$), 
 followed by a factor of two larger luminosities in all species of 
neutrinos at $t_{\rm pb} \ga 10$ ms for our model. This reflects a very approximate
 nature of our neutrino transport scheme. For example, $\alpha_\nu$ in Equation 
(\ref{eq:Qdiff1})), which regulates the neutrino diffusion timescale in our approximate 
 scheme, should change with time in reality and can be determined only by solving 
 a self-consistent neutrino transport. Concerning the RMS energy, our 1D-GR models 
  show also significantly higher energies (up to $\sim$ 30$\%$ enhancement) 
compared to the Boltzmann results (e.g., \cite{Sumiyoshi05,Bernhard10,Lentz11}).
Admitting that there is no doubt about  
 the importance of implementing a more detailed transport scheme in our GR simulations, 
 we think that our approximate neutrino transport is still useful for 
 the sake of this study, in which we explore to discuss possible impacts of GR 
 by comparing to the corresponding SR counterparts.

Top two panels in Figure \ref{pic:F3} compare the angle average of 
 the RMS neutrino energy for $\nu_e$ (left panel) and $\bar{\nu}_e$ (right 
 panel) after the break-out burst ($t_{\rm pb}
 \gtrsim 10$ ms). As seen, the RMS energies are highest
 for the 1D-GR model (black line), followed in order by 1D-SR, 3D-GR, and 3D-SR. 
In accord with 
 the previous 1D results \citep{Lentz11,Bernhard10,matthias04,Bruenn01}, our 3D results 
 (albeit limited to the early postbounce phase) 
support the expectation that the neutrino RMS energies 
increase when switching from SR to GR hydrodynamics.

\placefigure{pic:F3}

 The reason for the higher neutrino energy in GR models is that 
 the deeper gravitational well of GR produces more compact core structures, and thus
 hotter neutrino spheres at smaller radii. This is shown in the bottom 
 panels in Figure \ref{pic:F3} (compare the radii of the neutrino sphere between
 GR and SR models). The smaller neutrino energies for our 3D models
 compared to the corresponding 1D models (top panels) 
is due to their larger neutrino spheres (bottom panels). In our 3D models,
 the shock expands much further out assisted by convective overturns.
 (e.g., Figure \ref{shock}), which also extends the positions of the neutrino spheres.
  The enlargement of the neutrino sphere in multi-D models 
is qualitatively consistent with the 2D post-Newtonian results by 
\citet{Buras06a} including detailed neutrino transport.


As mentioned above, GR increases the neutrino luminosities and energies, while 
 the 3D hydrodynamics works to make the neutrino energy smaller.
 What we like to discuss finally is whether the gain effects of GR 
could or could not overcome the possible loss effects of GR that
 should shorten the residency time of material in the gain region.
 And multi-D effects join in the game because they could potentially 
 work against it to make the dwell time longer.


A widely prevailing indicator to diagnose the onset of the neutrino-driven explosions
 is the ratio of the residency timescale 
($\langle t_{\rm res} \rangle$) to the neutrino-heating timescale 
($\langle t_{\rm heat}\rangle $) in the gain region (e.g., \citet{Janka01,thomp05,Murphy08})\footnote{If this indicator is greater
 than unity, i.e. $\langle t_{\rm res} \rangle$/$\langle t_{\rm heat}\rangle > 1$,
 the neutrino heating proceeds fast enough to gravitationally unbind the fluid 
 element, otherwise the matter is swallowed by the neutrino-cooling layer}.
To estimate $\langle t_{\rm res} \rangle$, we employ the effective advection 
timescale (Equation (8) in \citet{Buras06a}), in which 
$\langle t_{\rm res} \rangle$ is determined by the crossing time of 
 mass shell between shock and gain radii. The local heating timescale
 is estimated 
 by the mass weighted average of the local heating timescale,
\begin{equation}
\label{eq:TauHeat}
\tau_{\rm heat}\equiv\frac{-\varepsilon_{\rm bind}}{\dot{Q}_{\nu,~{\rm total}}},
\end{equation}
 where we obey the Newtonian expression to estimate the local binding energy as
\begin{equation}
\label{eq:Ebind}
\varepsilon_{\rm bind}\equiv\rho\biggl(u^t\varepsilon+\frac{1}{2}v_iv^i+\phi_{NT}\biggr),
\end{equation}
 and the net neutrino heating rate is calculated by
 $\dot Q_{\nu,~{\rm total}}\equiv e^{6\phi}\alpha Q^\mu n_\mu$ (see, Equation (\ref{eq:GRenergy})). In estimating the heating timescale, the numerical cells that satisfy
 both $\varepsilon_{\rm bind}<0$ and $\dot Q>0$ are only taken into account.

\placefigure{pic:Res2Heat}

As seen from Figure \ref{pic:Res2Heat}, the shock revival seems most likely to 
 occur for the 3D-GR model (red line) in our simulation time, 
which is followed in order by 3D-SR, 1D-SR and 1D-GR models.
 Thanks to a more degree of freedom,
 the residency timescale becomes much longer for the 3D models than for the 1D models.
 In addition, the increase of the neutrino luminosity and 
RMS energies due to GR (Figure \ref{pic:F3}) enhances the timescale ratio 
up to the factor of $\sim$ 2 for 
 the 3D-GR model (red line) compared to the SR counterpart (blue line). 
 Therefore our results suggest that the combination of 3D and GR hydrodynamics 
 could provide the most favorable condition to trigger the neutrino-driven 
 explosions. 

As expected from Figure \ref{pic:Res2Heat}, the shock revival
 will never occur afterwards for the 1D models that have already shown the sign 
of a rapid shock recession. On the other hand, the curves for the 3D models
 stay constant for the last 30 ms before our simulation terminates. For the 
 15 $M_{\odot}$ progenitor employed in this paper, 
the neutrino-driven explosions are expected to 
 take place later than $\sim 200$ ms postbounce at the earliest \citep{bruenn}
 and it could be delayed after $\sim 600$ ms postbounce \citep{Marek09} as
 already mentioned.
 The parametric explosion models have shown that the earlier shock revival is good 
 for making the explosion energy larger (e.g., \citet{Nordhaus10}). The onset timescale of the neutrino-driven 
explosions predicted in 2D models \citep{Marek09,bruenn,Suwa10,Suwa11} 
could be shorter if the combination effects of GR and 3D would have been
 included. We anticipate that this can be a possible remedy to turn 
the relatively underpowered 2D explosions into the 
 powerful ones. To draw a robust conclusion, the energy and angle dependence of 
 the neutrino transport should be accurately incorporated in 
 our full GR simulations 
 with the use of more detailed set of weak interactions. This work 
 is only the very first step towards the climax to investigate these fascinating
 issues.

\section{Summary}
\label{sec:Summary}
We presented the results from the first-generation multi-D 
 core-collapse simulations in full GR that include 
 an approximate treatment of neutrino transport. 
 Using a M1 closure scheme with an analytic variable Eddington factor, 
 we solved the energy-independent set of 
radiation energy and momentum based on the Thorne's momentum formalism.
 To simplify the source terms of the transport equations, 
 a methodology of multiflavour neutrino leakage scheme was partly employed.
 Our newly developed code was designed to evolve the Einstein field
 equation together with the GR radiation hydrodynamic equations
 in a self-consistent manner while satisfying the Hamiltonian and momentum constraints. 
 An adaptive-mesh-refinement technique implemented in the three-dimensional (3D) 
 code enabled us to follow the dynamics starting from the onset of 
gravitational core-collapse of a 15 $M_{\odot}$ star, through bounce, 
up to about 100 ms postbounce in this study. By computing four models that differ by
  1D or 3D and by switching from SR to GR hydrodynamics, 
 we studied how the spacial multi-dimensionality and GR would affect the 
dynamics in the 
 early postbounce phase.
 Our 3D results support the anticipation in the previous 1D results that the neutrino 
luminosity and the
average neutrino energy of any neutrino flavor  in the postbounce phase generally
 increase 
 when switching from SR to GR hydrodynamics. 
This is because
 the deeper gravitational well of GR produces more compact core structures, and thus
 hotter neutrino spheres at smaller radii. 
By analyzing the residency to the neutrino-heating timescale in the gain regions,
 we pointed out that 
the criteria to initiate neutrino-driven explosions could be most easily
 satisfied in the 3D models, irrespective of the SR or GR hydrodynamics. 
 Keeping caveats in mind the omission of 
 energy- and angle-dependence of the radiation fields and the use of reduced set of 
 weak interactions in the present
 algorithm, our results indicated that the combination of 3D 
hydrodynamics and GR should provide the most favorable condition
  to drive a robust neutrino-driven explosion.
 On top of the omission of the spectral and angle dependence of
 the neutrino transport, we think the most urgent task is to replace 
 the leakage scheme with more realistic modeling of the source terms.

 In our 3D simulation, the numerical resolution behind the standing 
accretion shock is a few kilometers, 
which is not good enough to capture the growth of SASI accurately \citep{Sato09}. 
 The numerical viscosity is expected to be large especially in the vicinity of
 the shock, which may affect the growth of the SASI. It could also 
 affect the growth of the turbulence in the postshock convectively active regions,
 which is very important to determine the success or failure of the neutrino-driven 
 mechanism. To clearly see these effects of numerical viscosity, 
we need to conduct a convergence test in which a numerical gridding is changed 
in a parametric way (e.g. \citet{Hanke11}), although it is too computationally 
 expensive to do so for our 3D-GR models at present. 
An encouraging news is that we have an access to 
 the ``K-computer", which is the fastest one in the world as of November 2011. 
 Not in the distant future, we hope to report our 3D-GR models with much 
 higher resolutions to check the convergence of the present results.

The most up-to-date neutrino transport code in core-collapse supernova simulations 
 can treat the multi-energy and multi-angle 
transport in 2D \citep{ott_multi} and even in 3D simulations \citep{Sumi11} but 
mostly in the Newtonian hydrodynamics (see, however, \citet{bernhard11,Bernhard12}). 
 As was originally pointed out by \citet{Schwarz67} in the late 60's, our exploratory
 results also support the importance of GR to draw a 
robust conclusion to the supernova mechanism, indeed. The combined effects of GR and 
3D\footnote{It is worth mentioning that the MHD effects also remain to be studied
(\citet{kota04b,taki04,taki09,burr07,fogli_B,Kuroda10,martin11,taki_kota}, see also \citet{kota06} for collective references therein).} should affect not only the supernova dynamics, but also the observational multi-messenger signatures (e.g., \citet{kotake12} 
for a recent review)
, such as gravitational-waves (e.g., \citet{ewald11,kotake09a,kotake09b,kotake11a,ottprl}), neutrino emission (e.g.,
 \citet{icecube,marek09b,lund}),
 and explosive nucleosynthesis (e.g., \citet{fujimoto,friedel}).
Keeping our efforts to improve the caveats mentioned 
above, we are going to study these fascinating subjects one by one in the near future.

\appendix
\label{appA}

\section{Determination of Neutrino Spheres}
\label{sec:Optical depth}

As illustrated in Figure \ref{pic}, we have to calculate the neutrino optical-depth
 ($\tau_{\nu}$)
 to determine the position of the neutrino spheres.
 It can be done by solving the following differential equation (actually by 
 a matrix inversion) 
\begin{eqnarray}
\label{eq:OpticalDepth}
\frac{x^i D_i \tau_\nu}{r}=- \kappa_\nu,
\end{eqnarray}
with an appropriate boundary condition ($\tau_\nu|_{r\rightarrow\infty}=0$).
Here $\kappa_\nu$ represents the neutrino opacity of each species.
Then the neutrino sphere is determined where the optical-depth exceeds 2/3 ($\tau_{\nu} = 2/3$). For the matrix solver, we take the same one to solve the Poisson equation 
(\ref{eq:SelfGravity}).
 As shown in Table 1, we only include a reduced, but the most fundamental 
 set of weak interactions in the supernova cores, which consists of
 the charged-current interactions; $n\nu_e\leftrightarrow e^-p$, $p\bar{\nu}_e\leftrightarrow e^+n$, $\nu_e A\leftrightarrow e^-A'$ and scattering processes; $\nu p\leftrightarrow \nu p$, $\nu n\leftrightarrow \nu n$, $\nu A\leftrightarrow \nu A$. Note $\nu$
 in the scattering processes, represents all species of neutrinos 
($\nu_e,\bar\nu_e,\nu_x$).
 The opacity for electron, anti-electron, and heavy-lepton neutrinos 
  can be expressed as
\begin{eqnarray}
\kappa_{\nu_e}=\kappa_{a}(\nu_e n)+\kappa_{a}(\nu_e A)+\kappa_{s}(\nu_e n)+\kappa_{s}(\nu_e p)+\kappa_{s}(\nu_e A),
\end{eqnarray}
\begin{eqnarray}
\kappa_{\bar\nu_e}=\kappa_{a}(\bar\nu_e p)+\kappa_{s}(\bar\nu_e n)+\kappa_{s}(\bar\nu_e p)+\kappa_{s}(\bar\nu_e A),
\end{eqnarray}
and 
\begin{eqnarray}
\kappa_{\nu_e}=\kappa_{s}(\nu_x n)+\kappa_{s}(\nu_x p)+\kappa_{s}(\nu_x A),
\end{eqnarray}
 respectively (e.g., \citet{Ruffert96}).
Here the subindex $a$ and $s$ denote absorption and scattering processes, respectively.
 Detailed descriptions for the expressions of each opacity can be found 
in \cite{Bruenn85,Burrows06}.

Since we do not transfer the number density of neutrinos in the present scheme, 
we evaluate the emergent root-mean-squared (rms) energy of neutrinos 
$\varepsilon_{s_\nu,i}$ in the following way. Note here that the subscript $i = 
 \nu_e, \bar{\nu}_e$ denotes the neutrino species.
We first project the neutrino sphere defined in the cartesian grids to the 
spherical polar grids, which gives us the position of the neutrino sphere 
 expressed in the polar grids as $R_{\nu,i}(\theta,\phi)$.
Then we identify $\varepsilon_{s_\nu,i}$ with the energy of neutrinos
 at the neutrino sphere assuming that they stream freely outwards with 
possessing the information of the last scattering surface. Then 
 $\varepsilon_{s_\nu}$ at arbitrary point $(R,\theta,\phi)$ may be expressed as
\begin{eqnarray}
\label{eq:Estreaming}
\varepsilon_{s_\nu,i}(R,\theta,\phi)\equiv\varepsilon_{\nu,i}(R_{\nu,i}
(\theta,\phi),\theta,\phi).
\end{eqnarray}
Here $\varepsilon_{\nu,i}$ in the right-hand-side 
denotes the neutrino energy at $R_\nu(\theta,\phi)$, which is estimated by
\begin{equation}
\varepsilon_{\nu,i} = k_B T \frac{F_3(\eta_\nu,0)}
{F_2(\eta_\nu,0)},
\end{equation}
where $F_k$ is the Fermi-Dirac integral and $\eta_\nu=\mu_\nu/k_BT$ is the
  degeneracy parameter with $\mu_\nu$, $T$, $k_B$ representing the 
 neutrino chemical potential, matter temperature, and Boltzmann constant, respectively.

\subsection{Neutrino diffusion terms}
\label{sec:Neutrino diffusion terms}

Here we briefly summarize how to determine the diffusion term;
\begin{eqnarray}
\label{eq:Qdiff}
Q_{\nu,{\rm diff}}\equiv\int{g_\nu\frac{\varepsilon_\nu^k n_\nu}{\alpha_\nu T^{{\rm diff}}_\nu}d\varepsilon_\nu}
\end{eqnarray}
 according to \citet{Ruffert96,Rosswog03,Sekiguchi10}. In the above 
 equation, depending on $k$ (=1 or 0),  one can obtain the energy (or number)
 diffusion rate. $\alpha_\nu$ in Equation (\ref{eq:Qdiff}) and $\beta_\nu$ in Equations
 (\ref{eq:QmuC})-(\ref{eq:QmuH}) are model parameters that affect the  
 neutrino diffusion timescale and the position of the neutrino spheres, respectively.
 We adjust these parameters ($\alpha_\nu=2$ and $\beta_\nu=3/2$)
 in such a way to fit the neutrino luminosity obtained in the 1D results using 
  the IDSA scheme
(see Appendix \ref{sec:Tests for radiation part}). 
 $g_{\nu}$ ($g_{\nu_e}=g_{\bar\nu_e}=1$ and $g_{\nu_x}=4$) in 
 Equation (\ref{eq:Qdiff}) simply represents a multiplicity of neutrino species.  
 $n_\nu$ is the number density of neutrinos per each energy bin $\varepsilon_\nu$ in thermal equilibrium with matter and expressed by the Fermi-Dirac distribution function as
\begin{eqnarray}
\label{eq:dnde}
n_\nu=
\frac{4\pi}{(hc)^3}\frac{\varepsilon_\nu^2}{1+{\rm exp}(\frac{\varepsilon_\nu-\mu_\nu}{k_B T})},
\end{eqnarray}
here $\mu_\nu$ is the chemical potential of neutrinos.
We define the diffusion time scale of neutrinos as
\begin{eqnarray}
\label{eq:Tdiff}
T_\nu^{\rm diff}\equiv 3 \frac{\Delta x(\varepsilon_\nu)}{c}\tau(\varepsilon_\nu)
\end{eqnarray}
where $\Delta x(\varepsilon_\nu)$ is assumed to be $\Delta x(\varepsilon_\nu)=\tau(\varepsilon_\nu)/\kappa(\varepsilon_\nu)$.
We assume the optical depth and opacity can be expressed as
\begin{eqnarray}
\tau_\nu(\varepsilon_\nu)\sim \varepsilon_\nu^2\tilde\tau_\nu \nonumber\\
\kappa_\nu(\varepsilon_\nu)\sim \varepsilon_\nu^2\tilde\kappa_\nu \nonumber
\end{eqnarray}
by neglecting the higher order correction terms of $\varepsilon_\nu$.
$\tilde\tau_\nu$ and $\tilde\kappa_\nu$ are energy independent optical depth and opacity, respectively.
Finally, the energy integration in Eq.(\ref{eq:Qdiff}) is rewritten as
\begin{eqnarray}
\label{eq:Qdiff1}
Q_{\nu,{\rm diff}}&=&g_\nu\frac{1}{\alpha_\nu}\frac{4\pi c }{3(hc)^3}\frac{\tilde\kappa_\nu}{\tilde\tau_\nu^2}\int{\frac{\varepsilon_\nu^k}{1+{\rm exp}(\frac{\varepsilon_\nu-\mu_\nu}{k_B T})} d\varepsilon_\nu} \nonumber\\
&=&g_\nu\frac{1}{\alpha_\nu}\frac{4\pi c }{3(hc)^3}\frac{\tilde\kappa_\nu}{\tilde\tau_\nu^2}(k_B T)^k F_k (\eta_\nu,0)
\end{eqnarray}
where $\eta_\nu=\mu_\nu/k_BT$ is the degeneracy parameter of neutrino.

\label{sec:Appendix B}

\section{Implementation of EOS}
\label{sec:Implementation of EOS}
We employ Shen EOS \citep{Shen98} based on the Thomas-Fermi approximation
 and a minimization of the free energy within a relativistic mean field theory.
The available data\footnote{We use the updated version which is 
obtained from http://user.numazu-ct.ac.jp/~sumi/eos/\#shen2011} 
is tabulated as a function
 of the three thermodynamic variables of density, temperature, and electron fraction
 as  $(\rho, T, Y_{e})$. We smoothly interpolate/extrapolate the original data as;
$10^{3.1}\le\rho\le10^{16}\ {\rm g\ cm^{-3}}$ with 200 
equidistant intervals in a logarithmic scale, 
$10^{6}\le T \le10^{12}\ {\rm K}$ with 200 equidistant intervals in 
a logarithmic scale, and $0.01\le Y_{e} \le 0.55$ 
with 50 equidistant intervals in a linear scale.
The interpolation is performed first by a 
bicubic interpolation in the $\rho$-$T$ plane and then by a 
 cubic interpolation for $Y_e$ direction.

Thermodynamic variables such as the total pressure, internal energy, and entropy reads,
\begin{eqnarray}
P(\rho,T,Y_e)&=&P_b+P_{e^-}+P_{e^+}+P_\gamma, \\
\varepsilon(\rho,T,Y_e)&=&\varepsilon_b+\varepsilon_{e^-}+\varepsilon_{e^+}+\varepsilon_\gamma, \\
s(\rho,T,Y_e)&=&s_b+s_{e^-}+s_{e^+}+s_\gamma,
\end{eqnarray}
 where subscripts $b$, $e^-$, $e^{+}$ and $\gamma$
 denote the contributions from baryon, electron, positron and photon,
 respectively\footnote{Note that $\varepsilon_b$ does not include the atomic mass energy,
 which fits with the definition of Shen EOS (2011).}

\subsection{Supplement to the original Shen EOS}
 Since the original Shen EOS contains contributions only from baryons, we need to add 
the remaining contributions from leptons and photons. 
Although it can be done straightforwardly by 
 using formulae give in (e.g., \citet{Blinnikov}), we summarize them, for convenience,
 shortly in the following.

 To construct the leptonic EOS, we have only to determine 
the electron chemical potential $\mu_{e^{-}}$ 
from a given data-set of proton fraction, density, and temperature ($Y_p, \rho, T$).
This can be done by the charge neutrality condition $Y_p = Y_e$, 
 where $Y_e$ is defined by
\begin{eqnarray}
\label{eq:chargeneutrality}
Y_{e} = \frac{n_{e^{-}}-n_{e^{+}}}{n_b},
\end{eqnarray}
where $n_{e^{-/+}}$ and $n_b = \rho/m_u$ 
is the number density of electrons/positrons and bayrons with $m_u$ being the atomic
 mass unit.
   $n_{e^{-/+}}$ can be expressed by
\begin{eqnarray}
\label{eq:ne}
n_{e^{-/+}}=\sqrt{2}\frac{m_{e}c}{\hbar \pi^2}\beta^{3/2}\Bigl[F_{1/2}(\eta_{e^{-/+}},\beta)+\beta F_{3/2}(\eta_{e^{-/+}},\beta)\Bigr],
\end{eqnarray}
 where $\beta\equiv k_B T/m_e c^2$ and $\eta^{-/+}\equiv\mu_{e^{-/+}}/k_B T$ with 
 $k_{B}$, $m_e$, $T$, $\mu_{e^{-/+}}$ representing the Boltzmann constant, electron
 rest mass, temperature, and chemical potential of electron/positron, respectively
 \citep{Blinnikov}.
 $F_k(\eta,\ \beta)$ is the Fermi-Dirac integral,
\begin{eqnarray}
\label{eq:FDI}
F_k(\eta,\ \beta)=\int_0^{\infty}\frac{x^k(1+\beta x/2)^{1/2}}{e^{x-\eta}+1}dx,
\end{eqnarray}
 where the useful analytical formulae of the integral (with their derivatives) are given in \citet{Tooper69,Miralles96}.
 Remembering that $\eta^{-} = - \eta^{+}$ is satisfied in the supernova cores due to high temperature ($\ge
 10^{9}$ K),  one can find the solution of $\mu_{e^{-}}$ 
by Equations (\ref{eq:chargeneutrality}) and (\ref{eq:ne}). 
 Thus the total pressure, specific internal energy, entropy (per nucleon) can be
  readily calculated as
\begin{eqnarray}
\label{eq:PreE}
P_{e}=\frac{2\sqrt{2}}{3}\frac{m_e^4 c^5}{\hbar^3 \pi^2}\beta^{5/2}
\Bigl[F_{3/2}(\eta_e,\beta)+\frac{\beta}{2} F_{5/2}(\eta_e,\beta)\Bigr],
\end{eqnarray}
\begin{eqnarray}
\label{eq:EintE}
\varepsilon_{e}=\sqrt{2}\frac{m_{e}^2c^3}{\hbar \pi^2}\beta^{5/2}
\Bigl[F_{3/2}(\eta_e,\beta)+\beta F_{5/2}(\eta_e,\beta)\Bigr] \rho^{-1},
\end{eqnarray}
\begin{eqnarray}
\label{eq:EntropyE}
s_{e}=\Biggr[ \frac{\rho\varepsilon_{e}+P_{e}- n_{e}\mu_{e}}{\rho T N_A k_B}  \Biggl],
\end{eqnarray}
 where $N_A$ is the Avogadro constant. 

The contribution from photons is expressed as,
\begin{eqnarray}
\label{eq:PreGamma}
P_{\gamma}=\frac{1}{3}a_r T^4,
\end{eqnarray}
\begin{eqnarray}
\label{eq:EintGamma}
\varepsilon_{\gamma}=\frac{1}{\rho}a_r T^4,
\end{eqnarray}
\begin{eqnarray}
\label{eq:EntropyGamma}
s_{\gamma}=\frac{4a_r T^4}{3\rho T N_A k_B},
\end{eqnarray}
 where $a_r = 8 \pi^5 k^4/(15 c^3 h^3)$ denotes the radiation constant.


\subsection{Primitive recovery}
\label{sec:Primitive recovery}
Since we evolve hydrodynamic equations in a conservative form, we need
 to obtain primitive variables from the conservative ones.
For the primitive recovery, we first solve the following simultaneous equations to obtain $Z\equiv\rho hW^2$ and the Lorentz factor $W$ \citep{Cerda08,Kuroda10}
\begin{eqnarray}
(Z^2-S^2)W^2-Z^2=0\\
\tau+D-Z+P(Z,W,Y_e/Y_l^t)=0
\end{eqnarray}
for a given conservative set of 
variables $(\rho_\ast,S_i,\tau)$ and the electron/total lepton fraction $Y_e/Y_l^t$.
In the above equations, $S^2\equiv\gamma^{ij}S_iS_j$ and $D\equiv\rho_\ast/e^{6\phi}$.
$P$ is the pressure and can be determined once the enthalpy $h=Z/DW$ and the rest mass density $\rho=D/W$ are given.
We iteratively solve these equations by the Newton-Raphson method 
until the sufficient convergence is achieved.

\subsection{The sound velocity}
\label{sec:The sound velocity}
As given in \citet{Shibata05b}. the sound velocity is expressed as,
\begin{equation}
\label{eq:sound velocity}
c_s=\sqrt{\frac{1}{h}\left[ \frac{\partial P}{\partial \rho} \Biggr| _\varepsilon +\frac{P}{\rho^2}\frac{\partial P}{\partial \varepsilon} \Biggr| _\rho \right]},
\end{equation}
where $P$ and $\varepsilon$ include the sum of contributions from baryon, electron, 
 and photon.
 Regarding the partial derivatives of the thermodynamical variables, 
we take a finite differencing of Shen's EOS table for the baryonic part, 
meanwhile we use analytical formulae of the Fermi-Dirac 
integrals given in \cite{Miralles96} for the leptonic sector.

\section{Numerical Tests}
\label{appC}
\subsection{Core-Collapse tests with Shen EOS}
 We first present the 1D (in the same manner as 1D-SR/GR models by neglecting the non-radial matter velocity and momentum) core-collapse run without neutrinos 
to validate the implementation of Shen EOS instead of the phenomenological
 one taken in the original code \citep{Kuroda10}.
In the case of the adiabatic collapse,
 the so-called prompt explosion is expected to occur
 for the $15 M_{\odot}$ star \citep{WW95} as reported by \citet{Sumiyoshi04}
 in their 1D GR Lagrangian simulations using the same EOS.

 Figure \ref{pic:AppF1F2} shows the profiles of density (left panel) and radial velocity
 (right panel) between the GR (solid line) and SR (dashed line) model, respectively.
\placefigure{pic:AppF1F2}

As can be seen, the central density $\rho_c$ (left panel) and 
the infall velocity (right panel) becomes higher for the GR model,
which bounces at $\rho_c = 4.5 \times 10^{14}~{\rm g}~{\rm cm}^{-3}$ 
 with its inner-core baryon mass ($M_{\rm IC} =0.91 M_{\odot} $) being $\sim0.1 M_{\odot}$ smaller compared to the SR counterpart.
 After bounce, the prompt shock propagates through the entire iron core (left panel
 in Figure \ref{pic:AppF3}) for both of the models. For the GR model, the shock
 reaches at a radius of $1000$ km at $\sim20$ ms after bounce, with its explosion energy in 
 the range of 1 - 1.5 $\times 10^{51}$ erg (right panel
 in Figure \ref{pic:AppF3}), which is consistent with those 
obtained in \citet{Sumiyoshi04}. 
\placefigure{pic:AppF3}

As for the numerical accuracy, we monitor the violation of the average Hamiltonian constraint
$C_{\rm hm}$ and the Arnowitt-Deser-Misner mass (ADM mass) $M_{\rm ADM}$.
We adopt the following form for $C_{\rm hm}$ \citep{Shibata03} 
\begin{eqnarray}
  \label{eq:L1normHamiltonianCon}
  C_{\rm hm}\equiv\frac{1}{M_{\rm bar}}\int{\frac{\rho_\ast\mathcal{H}}
    {\left[\Bigl|\tilde D^i\tilde D_i e^{\phi}\Bigr|+\Bigl|\frac{e^{\phi}\tilde R}{8}\Bigr|+\Bigl|2\pi (S_0+E)e^{-\phi}\Bigr|+\Bigl|\frac{e^{5\phi}}{8}\left(\tilde A_{ij}\tilde A^{ij}-\frac{2}{3}K^2\right)\Bigr|\right]}dx^3}
\end{eqnarray}
where $M_{\rm bar}\equiv\int\rho_\ast dx^3$ is the proper rest mass and $\mathcal{H}$ is the 
Hamiltonian constraint,
\begin{eqnarray}
\label{eq:HamiltonianCon}
\mathcal{H}=\tilde D^i\tilde D_i e^{\phi}-\frac{e^{\phi}\tilde R}{8}+2\pi (S_0+E)e^{-\phi}+\frac{e^{5\phi}}{8}\left(\tilde A_{ij}\tilde A^{ij}-\frac{2}{3}K^2\right)=0.
\end{eqnarray}
$M_{\rm ADM}$ can be written as
\begin{eqnarray}
\label{eq:Madm}
M_{\rm ADM}=\int \left[(S_0+E)e^{-\phi}+\frac{e^{5\phi}}{16\pi}\left(\tilde A_{ij}\tilde A^{ij}-\frac{2}{3}K^2-\tilde{\gamma}^{ij}\tilde{R}_{ij}e^{-4\phi}\right)
\right]dx^3.
\end{eqnarray}
Every time after the number of the AMR blocks is increased
 by the AMR procedure and also after every restart of simulation,
 we enforce the Hamiltonian constraint by re-solving
 the following Poisson equation;
\begin{eqnarray}
\nabla^2_{\rm flat} \psi=\frac{\psi\tilde R}{8}-2\pi (S_0+E)\psi^{-1}
-\frac{\psi^5}{8}\left(\tilde A_{ij}\tilde A^{ij}-\frac{2}{3}K^2\right)
-f^{ij}\tilde D_i\tilde D_j \psi+\delta^{ij}\tilde\Gamma^k_{ij}\partial_k \psi,
\label{reenforce}
\end{eqnarray}
 until sufficient convergence is achieved. Here $\nabla^2_{\rm flat}$ is the Laplacian in flat space, $\psi\equiv e^\phi$ and $f^{ij}\equiv \tilde \gamma^{ij}-\delta^{ij}$.

Here we shortly comment on the side effect of this re-enforcement on the 
 gravitational-wave content of the spacetime. The resulting change in $\psi$ 
 is at most $\sim0.01\%$ near the central region. 
This means that the three metric $\gamma_{ij}(\propto (1 + \psi)^4)$ is also altered 
 at the level of $\sim0.01\%$ after the re-setting of the Hamiltonian constraint.
 For the sake of this study, this is negligibly small. On the other hand, when we 
 would deal with a much more massive progenitor, in which the spacetime could be 
 strongly curved especially in the case of black hole formation,
 we may have to be much more careful about the re-enforcement procudure.

In Figure \ref{error}, we plot violation of the Hamiltonian constraint 
$C_{\rm hm}$ ({\it dash}-{\it dotted}), baryon mass $M_{\rm bar}$ ({\it dashed})
 and the ADM mass ({\it solid}) for our 3D-GR model.
Since our hydrodynamic equations are in conservative forms, the baryon mass is
well conserved other than the inflowing materials through the outer computational boundary.
In regard to the ADM mass, even though it shows fluctuations especially after the core bounce,
the global trend shows similar behavior to $M_{\rm bar}$ and the the fluctuations are well below
1\% and we thus consider our numerical scheme preserves conservative variables
with sufficient accuracy.
Next, in regard to the constraint conditions, $C_{\rm hm}$ becomes larger in the postbounce phase, but
this is not surprising considering the complicated non-linear nature of the field equation and also the 
presence of the shock that makes the accuracy of the high order shock-capturing 
scheme down to the first-order scheme inevitably.
 $C_{\rm hm}$ is generally
kept less than $10^{-3}$ in the postbounce phase.

\placefigure{error}

In Figure \ref{error},  violation of the momentum constraint 
$C^i_{\rm mom}$;
\begin{eqnarray}
C^i_{\rm mom}\equiv \frac{1}{M_{\rm bar}} \int dx^3 \rho_\ast \frac
{\Bigl|\partial_j \tilde A^{ij}+\tilde\Gamma^i_{jk}\tilde A^{jk}+6\tilde A^{ij}\partial_j\phi-\frac{2}{3}\tilde\gamma^{ij}\partial_j K-8\pi\tilde\gamma^{ij} (S_j+F_j)\Bigr|}
{\Bigl|\partial_j \tilde A^{ij}\Bigr|+\Bigl|\tilde\Gamma^i_{jk}\tilde A^{jk}\Bigr|+\Bigl|6\tilde A^{ij}\partial_j\phi\Bigr|+\Bigl|\frac{2}{3}\tilde\gamma^{ij}\partial_j K\Bigr|+\Bigl|8\pi\tilde\gamma^{ij} (S_j+F_j)\Bigr|},
\end{eqnarray}
is also plotted for our 1D-GR model. Note only $x$ component, $C^x_{\rm mom}$, is shown
 since all other components of $C^i_{\rm mom}$ show almost the same profiles.
As denoted in the previous section, our manipulation of eliminating the non-radial 
components of fluid velocity could potentially 
violate the momentum constraint to a serious extent. 
On the other hand, the violation is shown to stay almost constant 
 with time in the postbounce phase (see, $C^x_{\rm mom}$ in Figure \ref{error}).
 Therefore we think that the very simple way 
to construct 1D models in 3D simulations that we propose in this work
would be quite useful.


\subsection{Tests for Transport Scheme}
\label{sec:Tests for radiation part}

For numerical tests of our transport algorithm, we present the check 
 for the trapped and streaming neutrinos, respectively.
 Note again that the sum of 
 the streaming and trapped neutrinos is transported by the evolution equations
 (Equations (\ref{rad1}), (\ref{rad2})). The streaming part 
 can be estimated by subtracting the trapped contribution from the sum,
 because the trapped part can be simply determined by local hydrodynamic 
 quantities (i.e. density, temperature, and $Y_e$).

Figure \ref{pic:AppF6} shows comparison of the RMS neutrino energy of the trapped 
 neutrino and the one obtained by the IDSA scheme below the neutrino sphere.
Since the IDSA can reproduce fundamental properties obtained 
in 1D full Boltzmann results \citep{idsa}, we think that the comparison with our 
approximate scheme with the IDSA is important.
This test is done for a given background of density, temperature and electron fraction 
profiles for several prebounce snapshots.

As seen, the assumption of $\beta$-equilibrium condition works 
well in the high density region and both of them show a quite similar profile there,
 in which neutrinos are essentially trapped by matter. 
The agreement regarding the position of the neutrino sphere also certificates 
our estimate of the RMS neutrino energy (Equation \ref{eq:Estreaming}).

\placefigure{pic:AppF6}

To check the properties of the streaming neutrinos, which is very relevant in the postbounce dynamics,
 we check the following two points, which are (1) whether the radiation energy flux 
falls with proportional to $r^{-2}$ above the neutrino sphere and (2)
whether the gain region, in which the net heating rate becomes positive, can be formed similar to previous 
studies. Figures \ref{pic:AppF7} shows radial profiles of the radiation (energy) flux that is obtained 
 by solving the closed set of the two-moments equations (Equations (\ref{rad1},\ref{rad2})).

\begin{figure}[htpb]
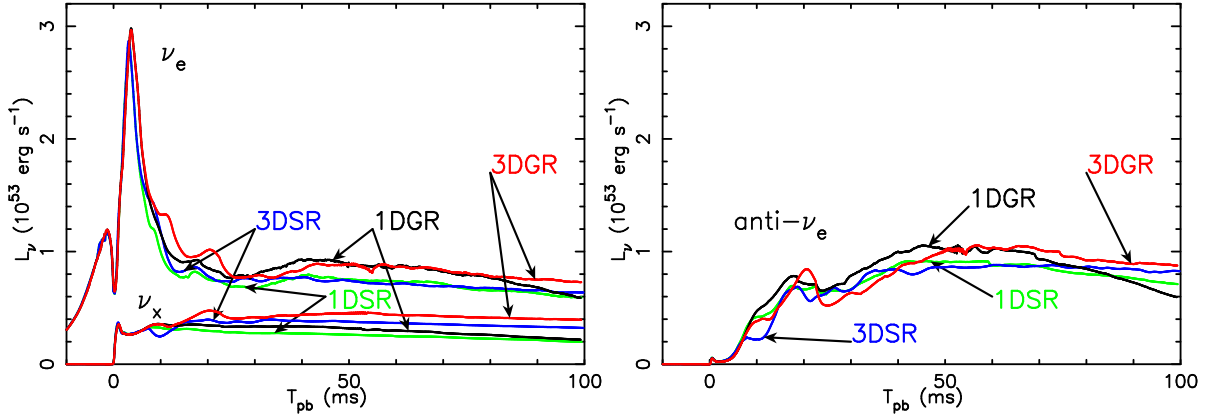

\begin{center}
\includegraphics[width=55mm,angle=-90.]{f13.eps}
\includegraphics[width=55mm,angle=-90.]{f14.eps}
\end{center}
  \caption{Neutrino luminosities of all neutrino flavors as a function of postbounce 
time (for $\nu_e$, $\nu_x$ (left panel), and for ${\bar{\nu}}_e$ (right panel),
 respectively.}
\label{lnu}
\end{figure}

\begin{figure}[htbp]
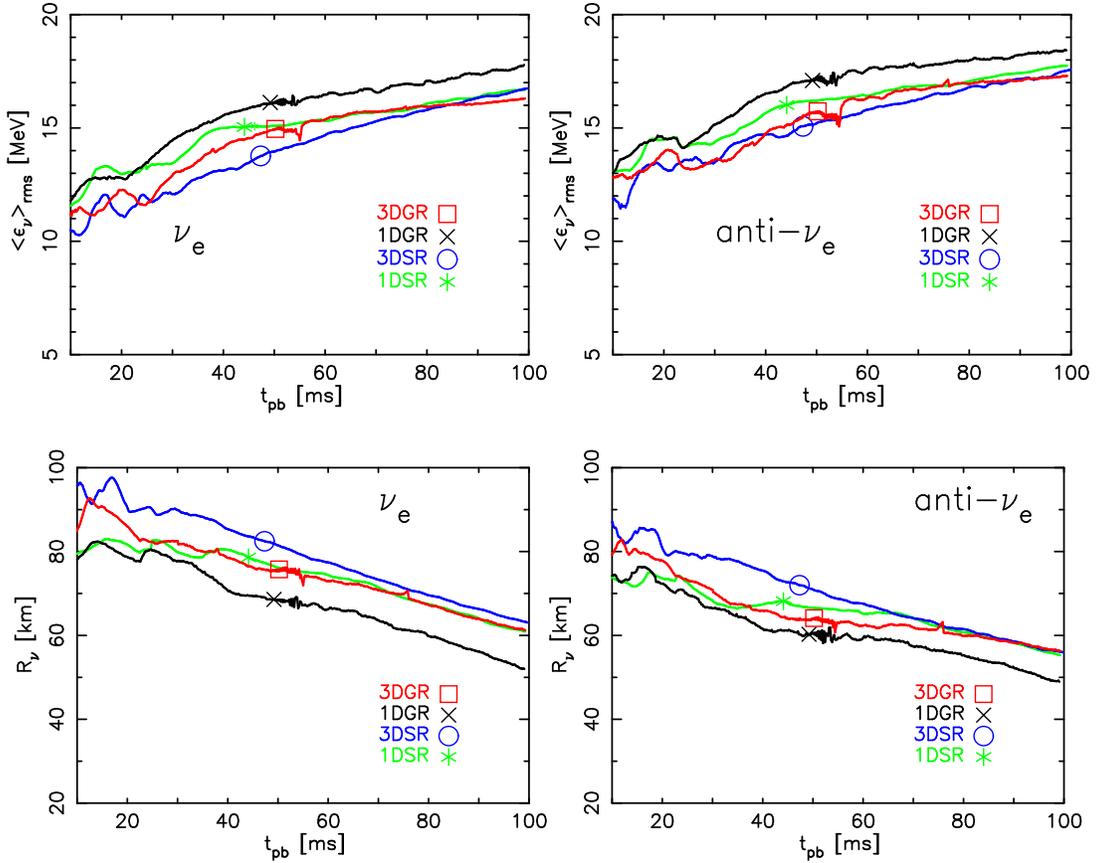

\begin{center}
\includegraphics[width=55mm,angle=-90.]{f15.eps}
\includegraphics[width=55mm,angle=-90.]{f16.eps}
\end{center}
\begin{center}
\includegraphics[width=55mm,angle=-90.]{f17.eps}
\includegraphics[width=55mm,angle=-90.]{f18.eps}
\end{center}
\caption{Evolution of the angle average RMS neutrino energy 
({\it upper} panels) and the radii of the neutrino spheres ({\it lower} panels) 
 for $\nu_e$ ({\it left}) and $\bar{\nu}_e$ ({\it right}). Colors are as in Figure 
 \ref{lnu}.}
\label{pic:F3}
\end{figure}

\begin{figure}[htpb]
\begin{center}
\includegraphics[width=65mm,angle=-90.]{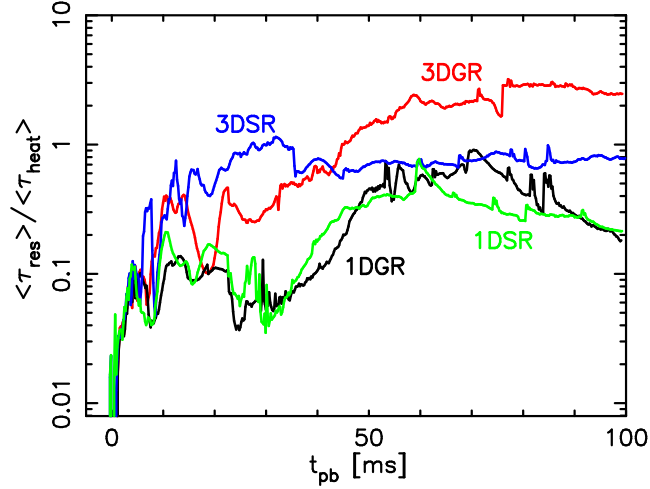}
\end{center}
  \caption{The ratio of the residency timescale to the heating timescale 
 for the set of our models as functions of post-bounce time (see text for the 
definition of the timescales).}
\label{pic:Res2Heat}
\end{figure}

\begin{figure}[htpb]
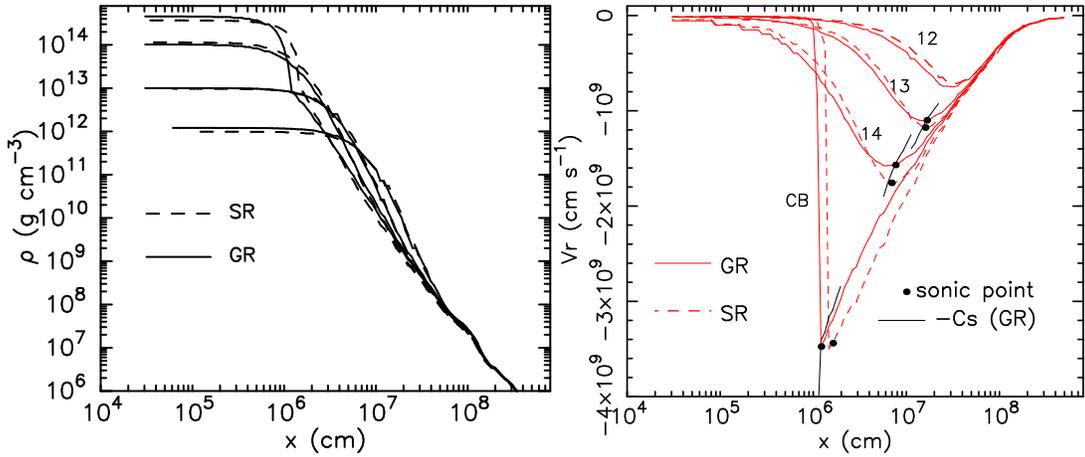

\begin{center}
\includegraphics[width=60mm,angle=-90.]{f20.eps}
\includegraphics[width=60mm,angle=-90.]{f21.eps}
\end{center}
  \caption{Profiles of the rest-mass density ({\it left}) and the radial velocity 
({\it right}) at times, when the central density reaches at $10^{12,13,14}$ g cm$^{-3}$ 
(from the bottom up to the top in the left panel, each density is 
denoted by $12, 13, 14$ and by ``CB" at bounce in the right panel). 
Solid and dashed line is for the GR and SR model, respectively. 
In the right panel, the profiles of the sound velocity and the sonic point
 are indicated by black solid lines and black points at the intersection between
 the radial velocity and the sound velocity.}
\label{pic:AppF1F2}
\end{figure}

\begin{figure}[htpb]
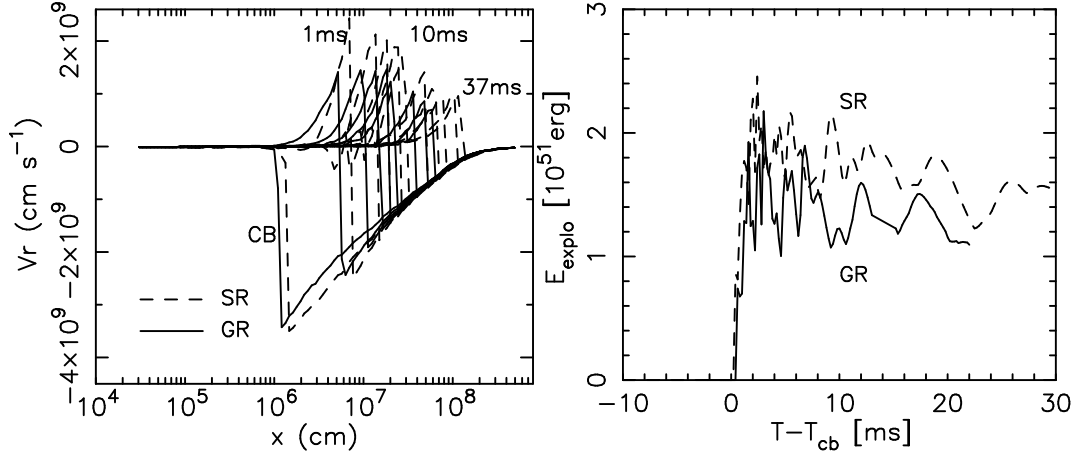

\begin{center}
\includegraphics[width=60mm,angle=-90.]{f22.eps}
\includegraphics[width=60mm,angle=-90.]{f23.eps}
\end{center}
  \caption{Evolution of the radial velocity (left panel) and the explosion energy 
(right panel) for the GR (solid line) and SR model (dashed line), respectively.
 ``CB" (core bounce) and the postbounce time is shown for reference. Note that 
 the explosion energy is defined in the Newonian limit that refers to the
integral of the energy over all zones that have a positive sum of the
specific internal, kinetic and gravitational energy. 
A smaller inner-core mass in the GR model leads to 
a smaller explosion energy because the amount of dissociation of iron nuclei 
becomes larger during the shock progagation.}
\label{pic:AppF3}
\end{figure}

\begin{figure}[htpb]
\begin{center}
\includegraphics[width=60mm,angle=-90.]{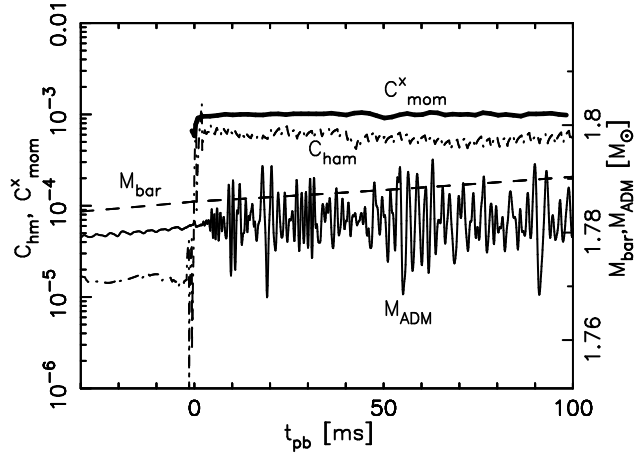}
\end{center}
\caption{Violation of the Hamiltonian constraint $C_{\rm hm}$ 
({\it dash-dotted}), baryon mass $M_{\rm bar}$ ({\it dashed})
and ADM mass ({\it thin-solid}) in our 3D-GR model are plotted against
the postbounce time. Note that the sudden decline in $C_{\rm ham}$ near 
 at bounce is due to the re-enforcement of the Hamiltonian constraint (e.g., Equation
 (\ref{reenforce})). 
We also plot violation of the Momentum constraint $C^x_{\rm mom}$ ({\it thick-solid}, 
here only $x$ component is plotted after the late collapse phase) for our 1D-GR model.}
\label{error}
\end{figure}

\begin{figure}[htpb]
\begin{center}
\includegraphics[width=60mm,angle=-90.]{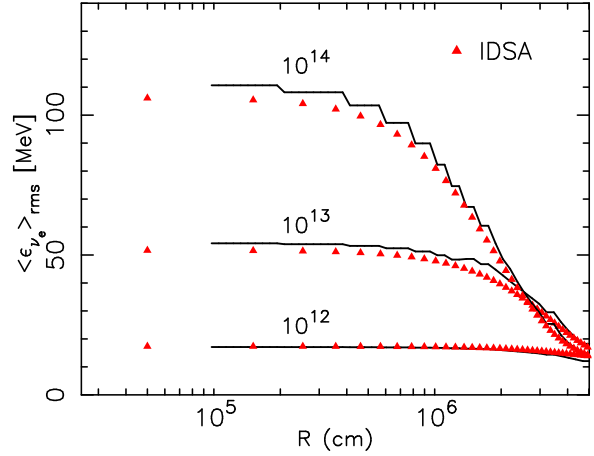}
\end{center}
  \caption{Profiles of the RMS neutrino energy at times, when the central density reaches $10^{12,13,14}$ g cm$^{-3}$
 obtained by the present scheme (black solid line) and by the IDSA scheme (red triangle), respectively.}
\label{pic:AppF6}
\end{figure}

\begin{figure}[htpb]
\begin{center}
\includegraphics[width=90mm,angle=-90.]{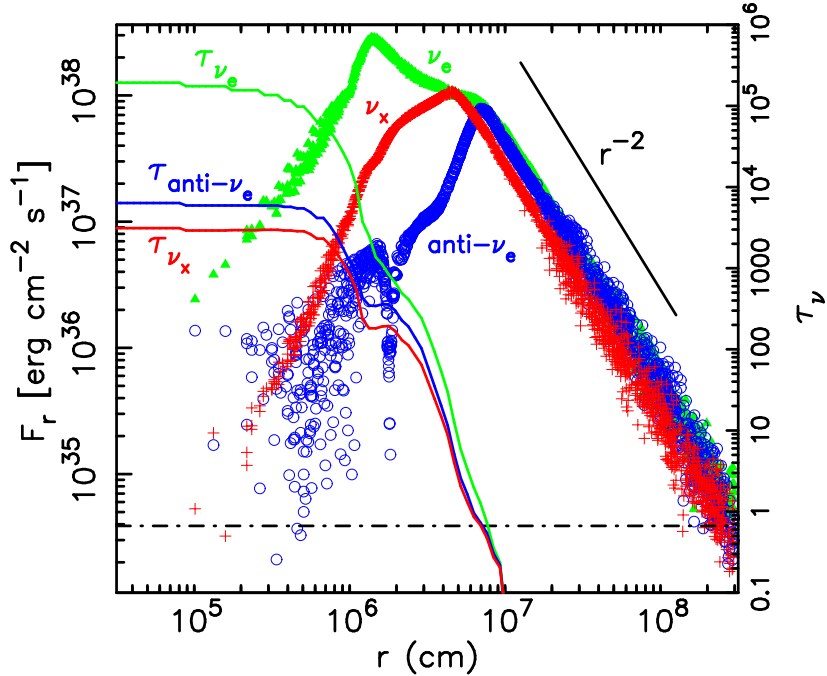}
\end{center}
\caption{Radial components of neutrino energy flux $F_{r,\nu}$ in all flavors are plotted by color coded points ({\it green} 
$\nu_e$, {\it blue} $\bar\nu_e$ and {\it red} $\nu_x$).
 The profiles of optical depth $\tau_{\nu}$ are shown by color-coded solid lines.
The horizontal dash-dotted line is drawn for $\tau=2/3$. As a reference, black solid line represents the slope of $r^{-2}$ in the log-scale. The data are 
  at 20ms after bounce in the 3D-GR model 
(see, section \ref{sec:Models and numerical methods}).}
\label{pic:AppF7}
\end{figure}

As seen, the energy fluxes change with $r^{-2}$ (compare with the black line) irrespective of neutrino 
 species outside the neutrino spheres. Note that the position of the neutrino spheres can be seen 
 as the intersection point between the horizontal line ($\tau = 2/3$) and the solid lines. The radii of 
 the neutrino spheres obeys a canonical order $R_{\nu_x} < R_{\bar{\nu}_e} < R_{\nu_e}$.
 The emergent neutrino energy flux can be estimated by $4\pi r^2$, $ 4\pi r^2 \times(F_{r,\nu_e},F_{r,\bar\nu_e},F_{r,\nu_x})\sim(9\times10^{52},8\times10^{52},4\times10^{52})$ erg s$^{-1}$, which are all in good agreement with the luminosity defined
by Equation (\ref{eq:Lnu}) as plotted in Figure \ref{lnu}.

\placefigure{pic:AppF7}

\begin{figure}[htpb]
\begin{center}
\includegraphics[width=90mm,angle=-90.]{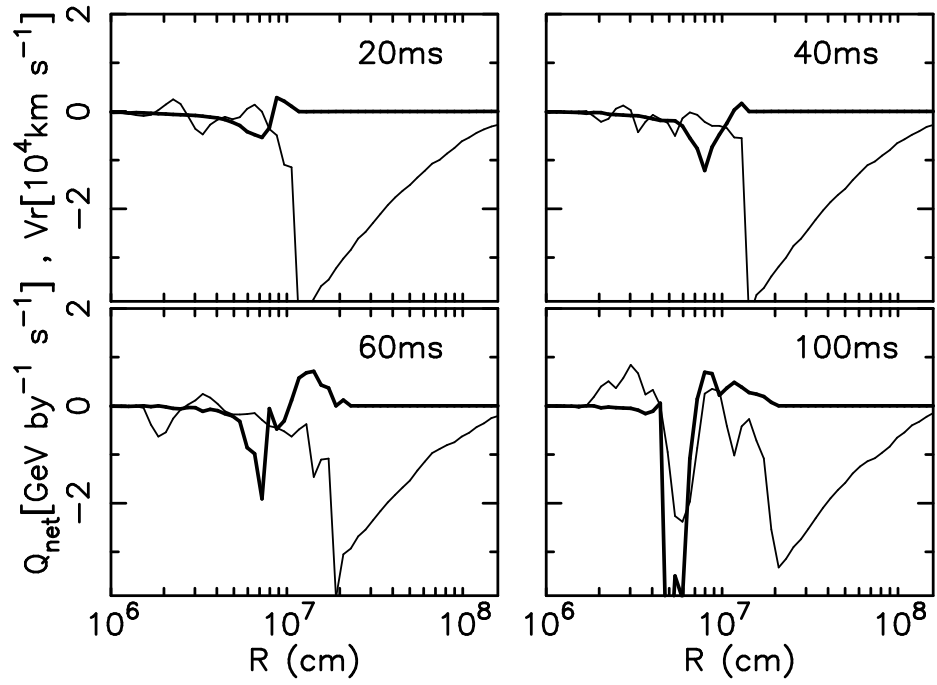}
\end{center}
\caption{Profiles of the net heating rate $Q_{\rm net} [{\rm GeV\ by^{-1}\ s^{-1}}]$ ({\it thick}) and the radial velocity $v_r$ 
({\it thin}) along the $x$ axis for our 3D-GR model at selected postbounce epochs. Note 
 that $v_r$ is normalized by $10^4$ km s$^{-1}$.}
\label{pic:F12}
\end{figure}

 Figure \ref{pic:F12} shows evolution of the net heating rate and 
radial velocity along the $x$ axis for our 3D-GR model 
(see, Sec. \ref{sec:Models and numerical methods}) at selected postbounce epochs.
 As the passive shock propagates (from top left to bottom right panels), the 
  gain region also gets larger. This reflects that the neutrino absorption on free 
nucleons predominantly takes place in the (enlarging) postshock region. The positive 
peak in the net heating rate is shown to be around $0.2 {\rm GeV}/{\rm nuc}/{\rm s}$ in the first 100 ms postbounce, which is in agreement with those in previous studies 
(e.g., \citet{Liebendorfer01,Sumi05b,ott_multi}).

\placefigure{pic:F12}

\acknowledgements{
 We are grateful to S.Yamada, K.Sumiyoshi, M.Liebend\"orfer, H.Nagakura, and Y.Suwa 
for stimulating discussions.
 TK is grateful to H.Umeda and T.Kajino for helpful exchanges and to M. Shibata,
 and Y. Sekiguchi for informative exchanges.
 KK and TT are thankful to 
K. Sato for continuing encouragements. 
Numerical computations were carried on in part on XT4 and 
general common use computer system at the center for Computational Astrophysics, CfCA, 
the National Astronomical Observatory of Japan, and also on
 SR16000 at YITP in Kyoto University.
 This 
study was supported in part by the Grants-in-Aid for the Scientific Research 
from the Ministry of Education, Science and Culture of Japan (Nos. 19540309, 20740150,
  23540323, and 23340069) and by HPCI Strategic Program of Japanese MEXT}

\newpage
\newpage

\newpage

\end{document}